%% file: main.tex
\documentclass[aps,prd,reprint,superscriptaddress,longbibliography]{revtex4-1}

\usepackage{graphicx,amssymb,amsmath,xcolor,graphicx, color}
\usepackage[normalem]{ulem}

\begin{document}


\title{Mass composition of ultra-high energy cosmic rays from distribution of their arrival directions with the Telescope Array}





\input{TA-author-20220427-revtex.tex}

\begin{abstract}
We use a new method to estimate the injected mass composition of ultrahigh cosmic rays (UHECRs) at energies higher than $10$~EeV.
The method is based on comparison of the energy-dependent distribution of cosmic ray arrival directions
as measured by the Telescope Array experiment (TA) with that calculated in a given putative model
of UHECR under the assumption that sources trace the large-scale structure (LSS) of the Universe.
As we report in the companion letter, the TA data show large deflections with respect to the LSS
which can be explained, assuming small extra-galactic magnetic fields (EGMF),
by an intermediate composition changing to a heavy one (iron) in the highest energy bin. 
Here we show that these results are robust to uncertainties in UHECR injection spectra,
the energy scale of the experiment and galactic magnetic fields (GMF). The assumption of weak EGMF,
however, strongly affects this interpretation at all but the highest energies $E > 100$~EeV,
where the remarkable isotropy of the data implies a heavy injected composition even in the case of strong EGMF.
This result also holds if UHECR sources are as rare as 
$2 \times 10^{-5}$~Mpc$^{-3}$, that is the conservative lower limit for the source number density.

\end{abstract}


\maketitle


\section{Introduction}
Ultra-high energy cosmic rays (UHECR) are charged particles of high energies $E > 1$~EeV ($1\, {\rm EeV}  = 10^{18}\, {\rm eV}$) that are reaching the Earth from space.
The UHECR spectrum is showing a steep decline at highest energies~\cite{HiRes:2007lra, PierreAuger:2008rol}, indicating some specific physical process. The nature of this steepening is related to UHECR mass composition at these energies --- the type of particles constituting the UHECR flux.
If the cutoff in the {\it injected} spectrum is high enough, the
steepening of the {\it observed} spectrum is associated with the Greisen-Zatsepin-Kuzmin (GZK) process --- the scattering of primary UHECR on the cosmic background radiation~\cite{Greisen:1966jv,Zatsepin:1966jv}. In this case the observed flux is enriched by protons, either primary or secondary.
At the same time, a lower injection cutoff manifests itself in the observed spectrum directly. In this case the flux at high energies consists of the same nuclei that were injected in sources~\cite{Aloisio:2009sj}.
Therefore, by estimating the UHECR mass composition at highest energies one could discriminate between these two scenarios.
However, this is a challenging task for standard UHECR measurement techniques.

The flux of UHECR is tiny, of order $1~{\rm km^{-2} sr^{-1} yr^{-1}}$ at $E \gtrsim 1$~EeV
and as small as $\sim 10^{-2}~{\rm km^{-2} sr^{-1} yr^{-1}}$ at highest energies of $E \gtrsim 100$~EeV.
Therefore, they can be detected only indirectly via extensive air showers (EAS) of secondary particles they produce in Earth's atmosphere.
The standard technique of the mass composition measurement employs the fluorescence detectors (FD) that are observing 
the ultraviolet light that EASs emit while propagating through the atmosphere.
Extracting the distribution of atmospheric depths of shower maxima ($X_{max}$) from the FD data and fitting it with simulated EAS of various primary particles one can estimate the {\it observed} UHECR mass composition~\cite{PierreAuger:2014sui, PierreAuger:2014gko, Abbasi:2014sfa, TelescopeArray:2018xyi}.
Being the most reliable mass composition measurement technique up to date, this method is still prone to uncertainties of high-energy hadronic models. 
In addition, FD measurements are possible only in moonless nights, which reduces the initially small UHECR statistics at the highest energies to $\sim 10\%$ of its full value. As a result, the FD measurements do not cover the physically most interesting region of highest energies.
In composition measurements with surface detectors (SD) the uncertainty due to high-energy hadronic interaction models
is either 
inherited from the FD by the cross-calibration~\cite{PierreAuger:2017tlx} or follows directly from the SD Monte-Carlo (MC)~\cite{Abbasi:2018wlq}, but in both cases the results are less accurate than those of the FD.  
There are also interesting proposals of mass composition reconstruction using neural networks~\cite{PierreAuger:2021fkf, Kalashev:2021vop}, but the fundamental problem of hadronic model dependence is not yet solved in this approach either.
Finally, the Pierre Auger observatory is now undergoing the surface detector upgrade that would allow it to measure electromagnetic and muonic parts of showers separately~\cite{Stasielak:2021hjm}. These measurements are expected to improve the composition-related 
discriminating power of the surface detector observations.

An alternative idea to use the UHECR anisotropy as a measure of their charge and hence mass composition has been proposed in Ref.~\cite{Kuznetsov:2020hso}.
There is a number of studies on the measurement of the UHECR anisotropy~\cite{PierreAuger:2017pzq, PierreAuger:2018zqu, PierreAuger:2018qvk, PierreAuger:2022axr, TelescopeArray:2014tsd, TelescopeArray:2020acv}, as well as several theoretical approaches that are using these measurements to unveil UHECR sources and mass composition~\cite{diMatteo:2017dtg, Allard:2021ioh, Ding:2021emg, Allard:2023uuk, Higuchi:2022xiv, Kuznetsov:2023jfw, Bister:2023icg}.
Our method has an advantage that it uses only the most robust UHECR observables: arrival directions and energies.
Comparing the energy-dependent distribution of UHECR arrival directions over the sky with the distribution expected in a generic model of sources with a given injected composition one can constrain this composition from the data.
The key ingredient of the method is the test statistics (TS) that summarizes the information contained in the 
arrival directions of the given event set in a single number: the mean deflection of the events from the sources that are assumed to follow the Large Scale Structure of the Universe. Due to shrinking of the attenuation horizon and decrease of magnetic deflections,  at highest energies the UHECR flux is expected to consist of isolated sources with different degrees of smearing for different primaries. This potentially allows one to constrain composition even at highest energies where the experimental statistics is small. The method is applied to the TA data with $E > 10$~EeV in the companion letter~\cite{TelescopeArray:2024oux}.
From the physical point of view, the most interesting result is the indication of a heavy mass composition at energies higher than 100~EeV.

In this paper we focus on the impact of various uncertainties that affect the compatibility of the composition models with the data: parameters of injected UHECR spectra, systematics of the energy scale, uncertainties of galactic and extragalactic magnetic fields, effect of the small number density of sources.
We show that most of these uncertainties have a negligible impact on the physical result mentioned above.

The paper is organized as follows: in Sec.~\ref{sec:experiment} we briefly introduce the Telescope Array experiment, the reconstruction procedure, and the data set used. In Sec.~\ref{sec:analysis} we describe the analysis method used in this study and give the details of the simulation of the mock UHECR sets. 
In Sec.~\ref{sec:results} we present the resulting constraints on composition models from the TA data. In Sec.~\ref{sec:uncert} we evaluate the impact of various uncertainties on these results. Sec.~\ref{sec:discussion} contains concluding remarks.

\section{Experiment, data and reconstruction}
\label{sec:experiment}
Telescope Array~\cite{TelescopeArray:2012uws, Tokuno:2012mi} is the largest cosmic-ray experiment in the Northern Hemisphere.
It is capable to detect EAS in the atmosphere initiated by cosmic particles of EeV energies and higher.
The experiment is located at $39.3^\circ$~N, $112.9^\circ$~W in Utah, USA, and has operated in a hybrid mode since May 2008.
It includes the  surface detector and 38 fluorescence telescopes grouped into three stations.
The SD consists of 507 scintillator stations of 3 m$^2$ each, placed
in a square grid with the 1.2~km spacing and covers the area of $\sim 700 {\rm km}^2$. The duty cycle of the SD is about 95\%~\cite{AbuZayyad:2012ru}. 

We use the standard TA SD reconstruction procedure as described in Refs.~\cite{AbuZayyad:2012ru, 2014arXiv1403.0644T}.
Each event is reconstructed by separate fits of shower geometry and lateral distribution function (LDF), which 
allows one to determine the shower arrival direction, core location and signal density at the distance 800~m from the core $S_{800}$.
The latter quantity together with the zenith angle is used to reconstruct the primary energy by making use of 
lookup tables derived from a full Monte-Carlo of EASs and the detector response~\cite{2014arXiv1403.0644T}.
Finally, the energy is rescaled by a correction factor $1/1.27$ to match the energy scale of the calorimetric TA FD technique.
The resolution of the arrival direction reconstruction is estimated as $1.5^\circ$ at $E \ge 10$~EeV~\cite{AbuZayyad:2012hv}.
The energy resolution is found to be $18\%$ in terms of logarithm of reconstructed to thrown energies ratio $\ln (E_{\rm rec}/E_{\rm MC})$ for $E_{\rm MC} \ge 10$~EeV~\cite{AbuZayyad:2012ru, 2014arXiv1403.0644T}.
The systematic uncertainty of the energy scale coming from FD is estimated to be 
$21\%$~\cite{TheTelescopeArray:2015mgw}.

To insure proper reconstruction of the primary particle parameters the following 
quality cuts are imposed~\cite{TelescopeArray:2014ahm}:
\begin{itemize}
\item  $E \ge 10$~EeV;
\item zenith angle $\le 55^\circ$;
\item number of ``good'' detectors in the fit $\ge 5$;
\item $\chi^2/{\rm d.o.f.} \le 4$ for both geometry and LDF fits;
\item pointing direction error $\le 5^\circ$;
\item $\sigma_{S_{800}}/S_{800} \le 0.25$;
\item detector with the largest signal is surrounded by 4 working detectors, there must be one working detector to the left, right, down, up on the grid of the largest signal detector but they do not have to be immediate neighbors of the largest signal detector.
\end{itemize}
These are the standard TA cuts used for anisotropy studies. In addition, we also eliminate the events induced by lightnings that mimic the EAS~\cite{ABBASI20172565, Abbasi:2017muv}. The lightning events are taken from the Vaisala lightning database compiled by the U.S. National Lightning Detection Network (NLDN)~\cite{NLDN}. We correlate the list of the NLDN lightning events detected within
15 miles from the Central Laser Facility of the TA during the full time of TA operation with the list of TA events. We remove all the TA events that occur within 10 minutes before or
after the NLDN lightnings. This cut was shown to reduce the total exposure by less then 1\%~\cite{TelescopeArray:2020hey}.

In the present study we use the TA SD data set obtained
during 14 years of operation from May 11, 2008 to May 10, 2022. The total number of events passing all the cuts is 5978; 19 of these events have energies larger than 100~EeV, including the highest energy event with $E = 244$~EeV~\cite{TelescopeArray:2023sbd}.

\section{Analysis}
\label{sec:analysis}
Our analysis closely follows that of Ref.~\cite{Kuznetsov:2020hso}. It is based on the computation and comparison of the same, properly defined test statistics (TS) for both UHECR data set and mock sets simulated in the assumption of various injected compositions.
The general outline of the method is the following. 
In general, each composition model is characterized by the fractions of injected species, spectral slopes and cut-off energies for each species. These parameters are not independent as the model has to reproduce the observed UHECR spectrum. In this analysis we limit ourselves to a simplified set of models in which {\em each species independently} is injected in such a way (see the details below) as to reproduce the observed spectrum in the energy range of interest. In this case the fractions of injected species are independent parameters --- those which we aim to constrain. For given fractions we generate a large mock set of UHECR events 
distributed according to flux maps computed for this composition with full account of attenuation and propagation effects.
The sources are assumed to trace the Large Scale Structure of the Universe. All other parameters affecting the UHECR flux distribution are fixed by some conservative assumptions as will be discussed below. Second, we define the test statistics that quantifies only the overall magnitude of the deflections of a given event set with respect to the LSS. This TS only involves parameters that are most robustly measured by the experiment: the event arrival directions and energies.
At the third step, we compute this TS for each mock event set and for the actual TA data set, and quantify the compatibility of each composition model with the data by means of the likelihood method. Finally, we estimate the impact of uncertainties of other parameters affecting the UHECR flux: shapes of injection spectra, galactic and extragalactic magnetic fields,  energy scale of the experiment and UHECR source number density. Varying these parameters in their experimentally allowed ranges we estimate how robust our conclusions about the composition are.

\subsection{Simulation of mock event sets}
\label{sec:analysis:sim}
\begin{figure*}
 \includegraphics[width=0.99\columnwidth]{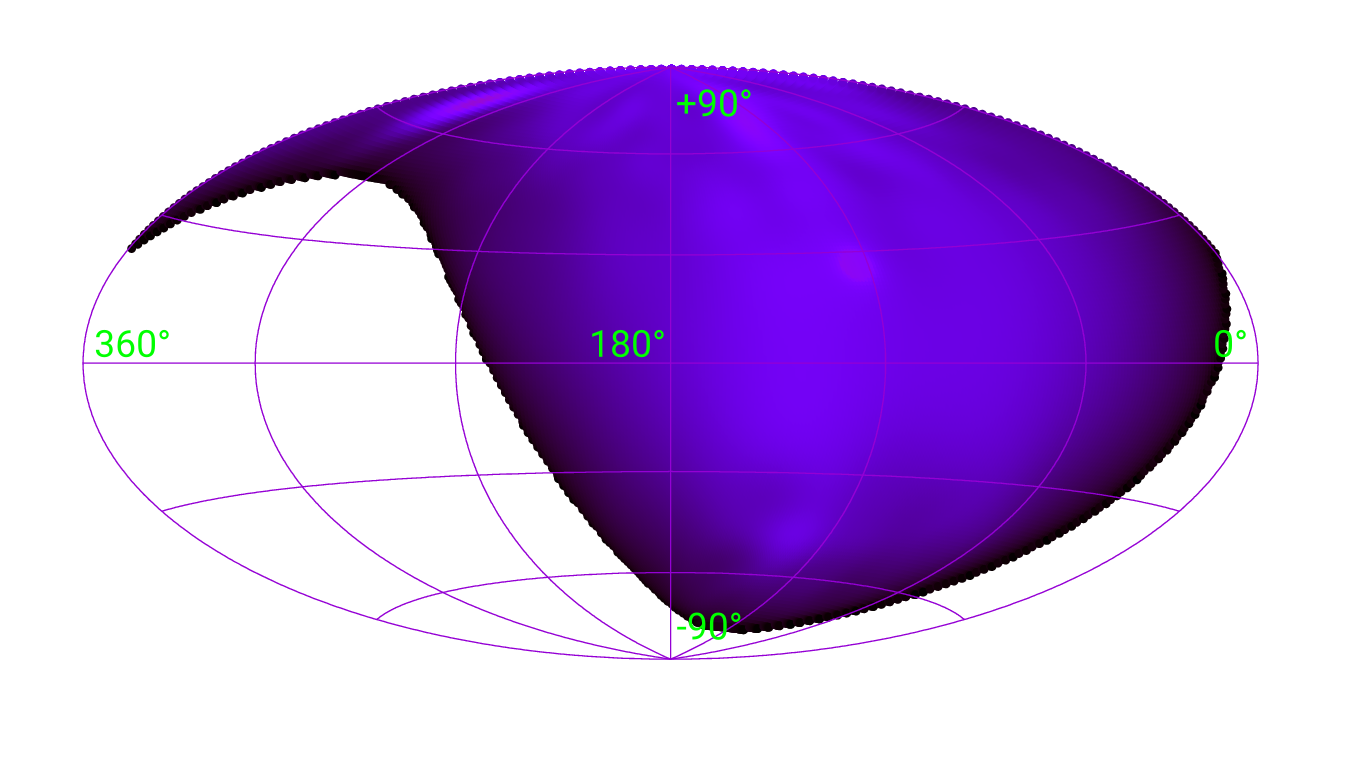}
 \includegraphics[width=0.99\columnwidth]{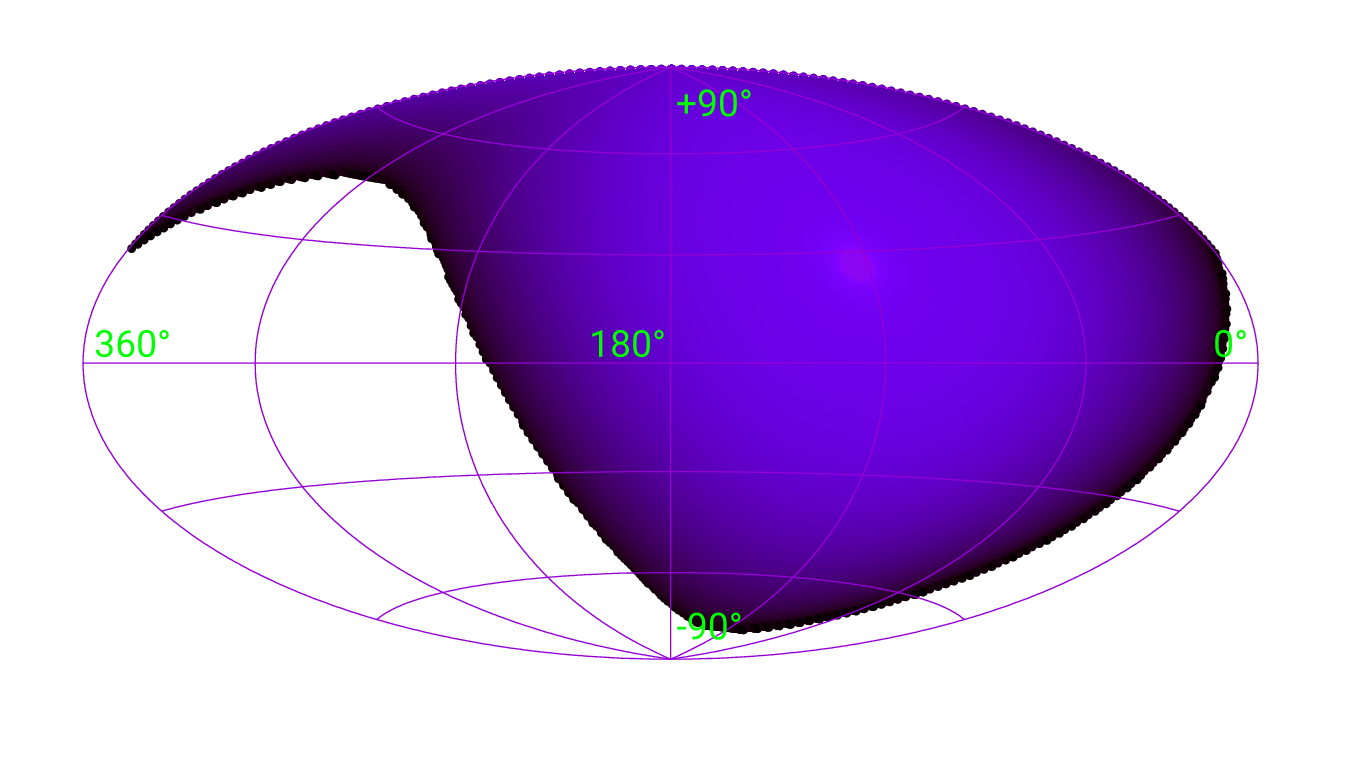}
 \includegraphics[width=0.99\columnwidth]{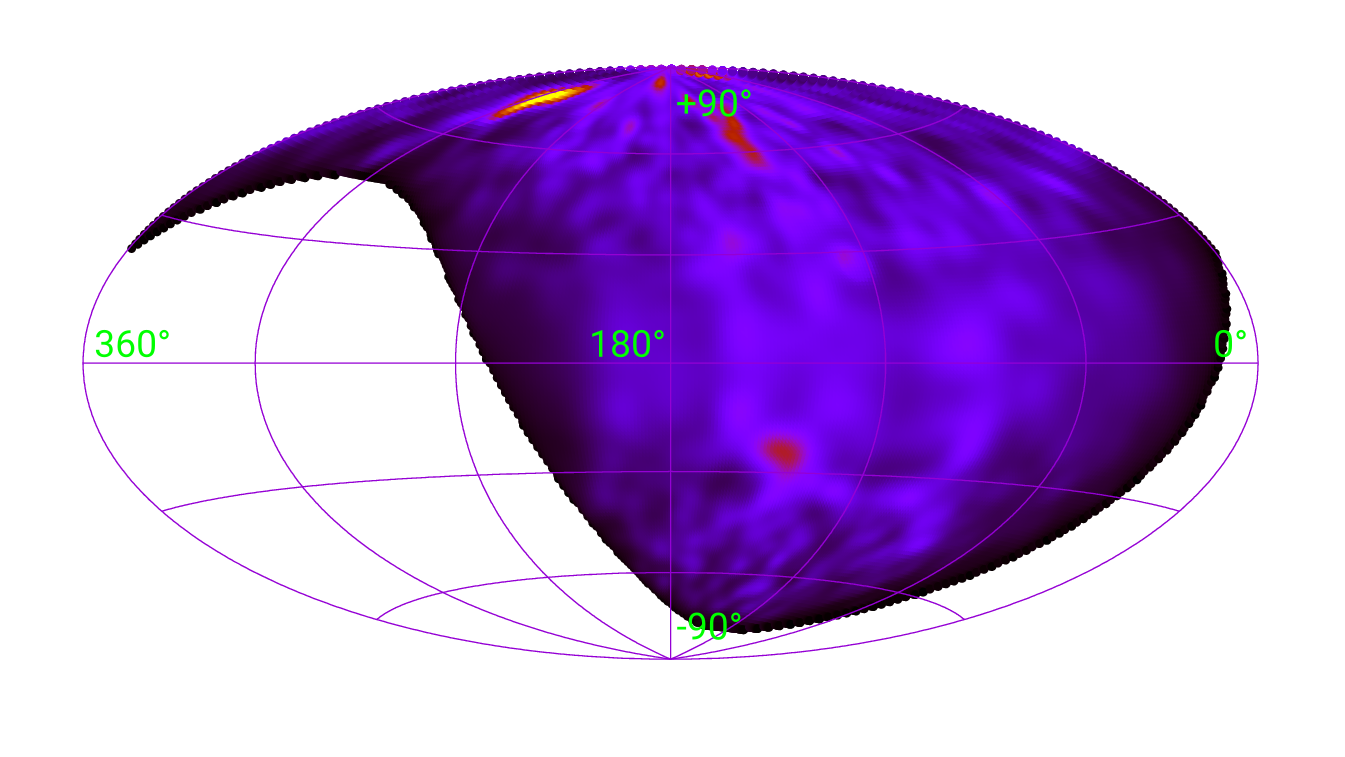}
 \includegraphics[width=0.99\columnwidth]{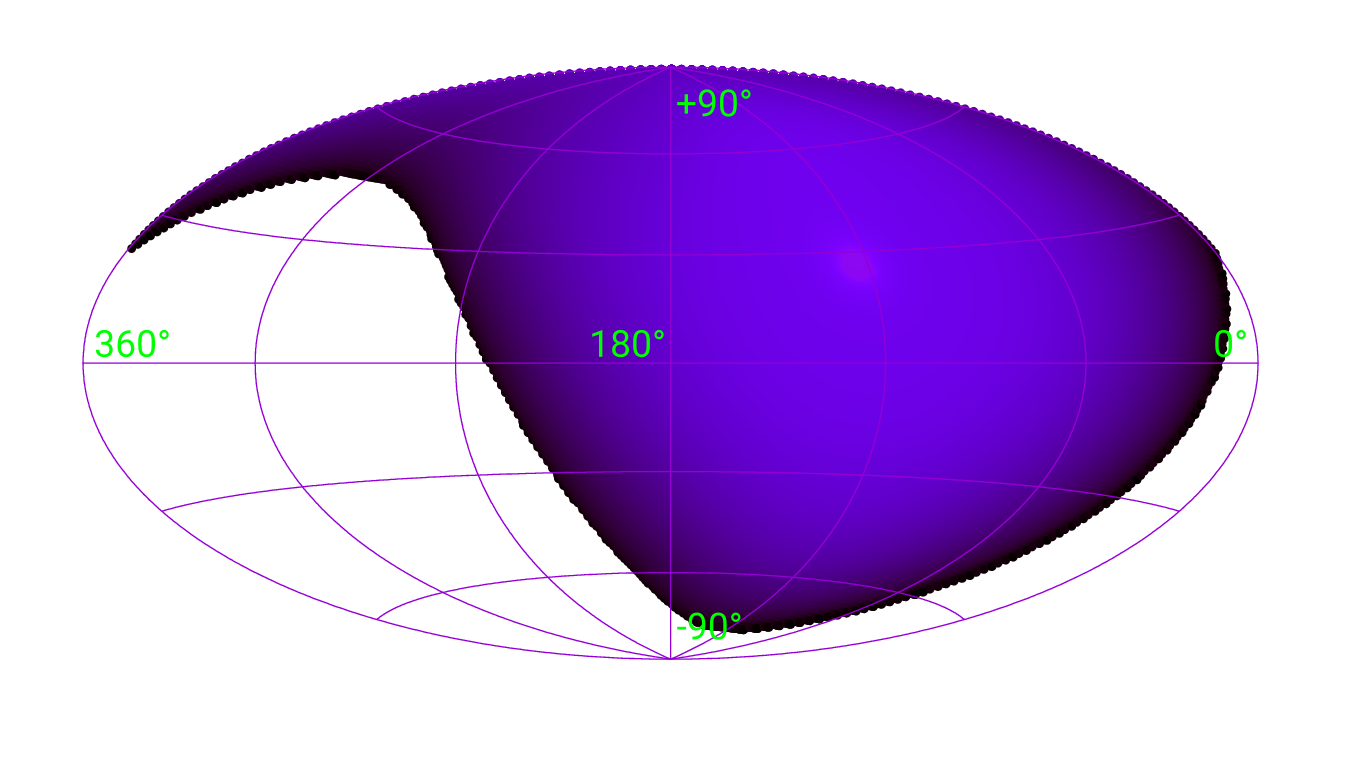}
 \includegraphics[width=0.99\columnwidth]{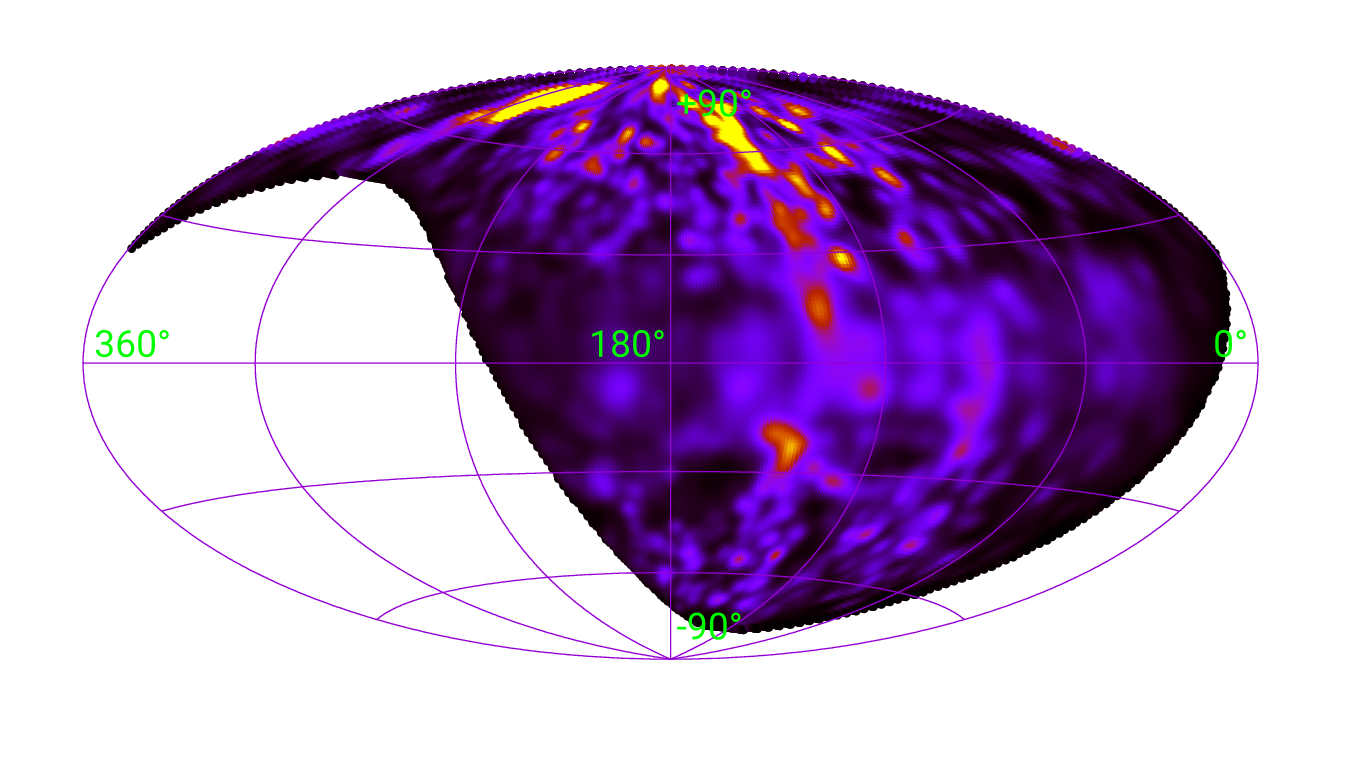}
 \includegraphics[width=0.99\columnwidth]{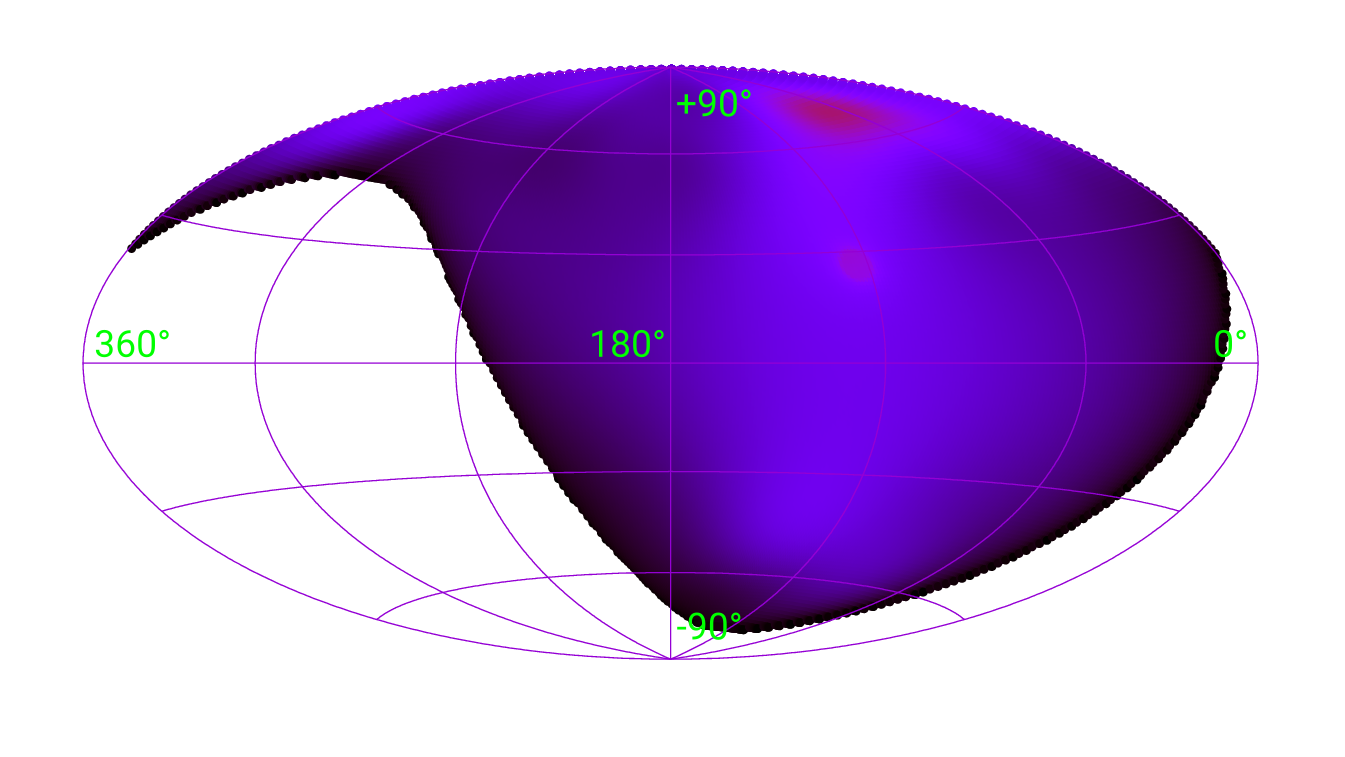}
 \includegraphics[width=0.99\columnwidth]{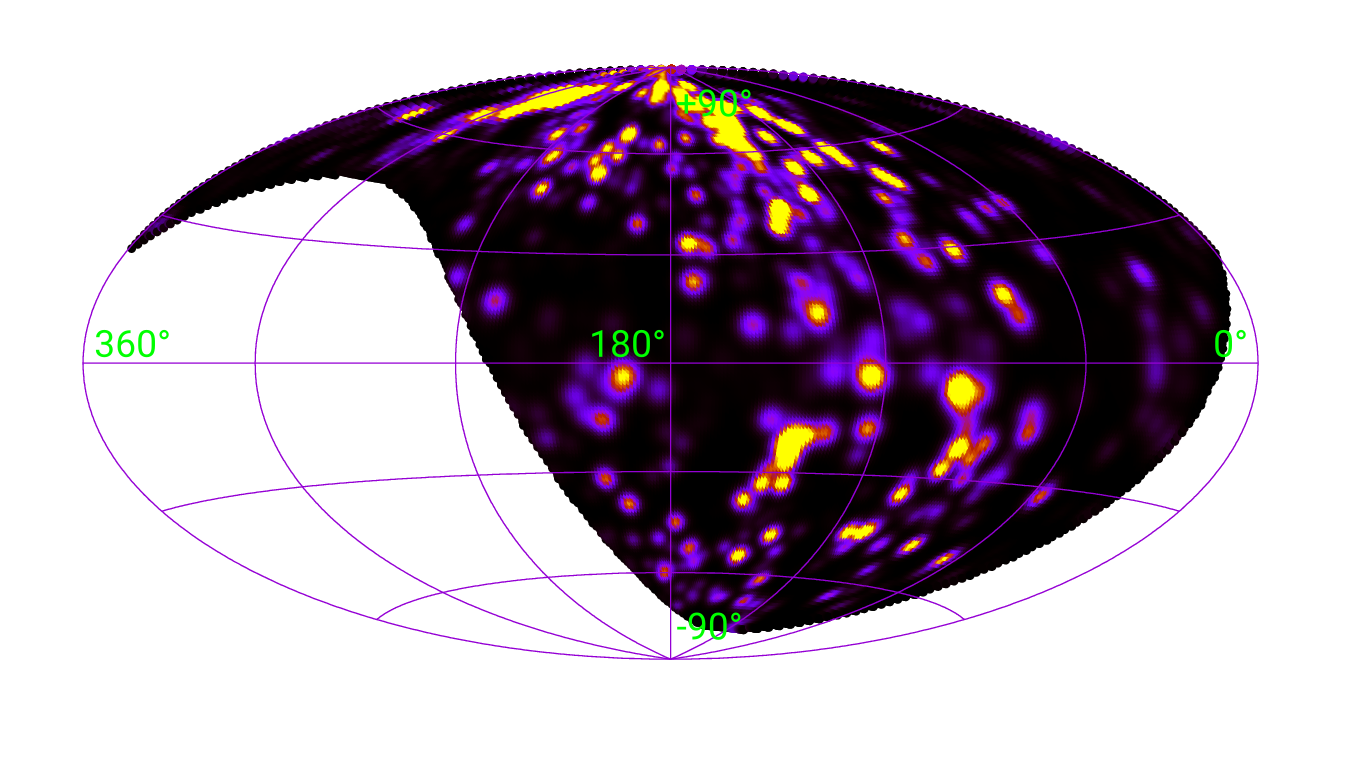}
 \includegraphics[width=0.99\columnwidth]{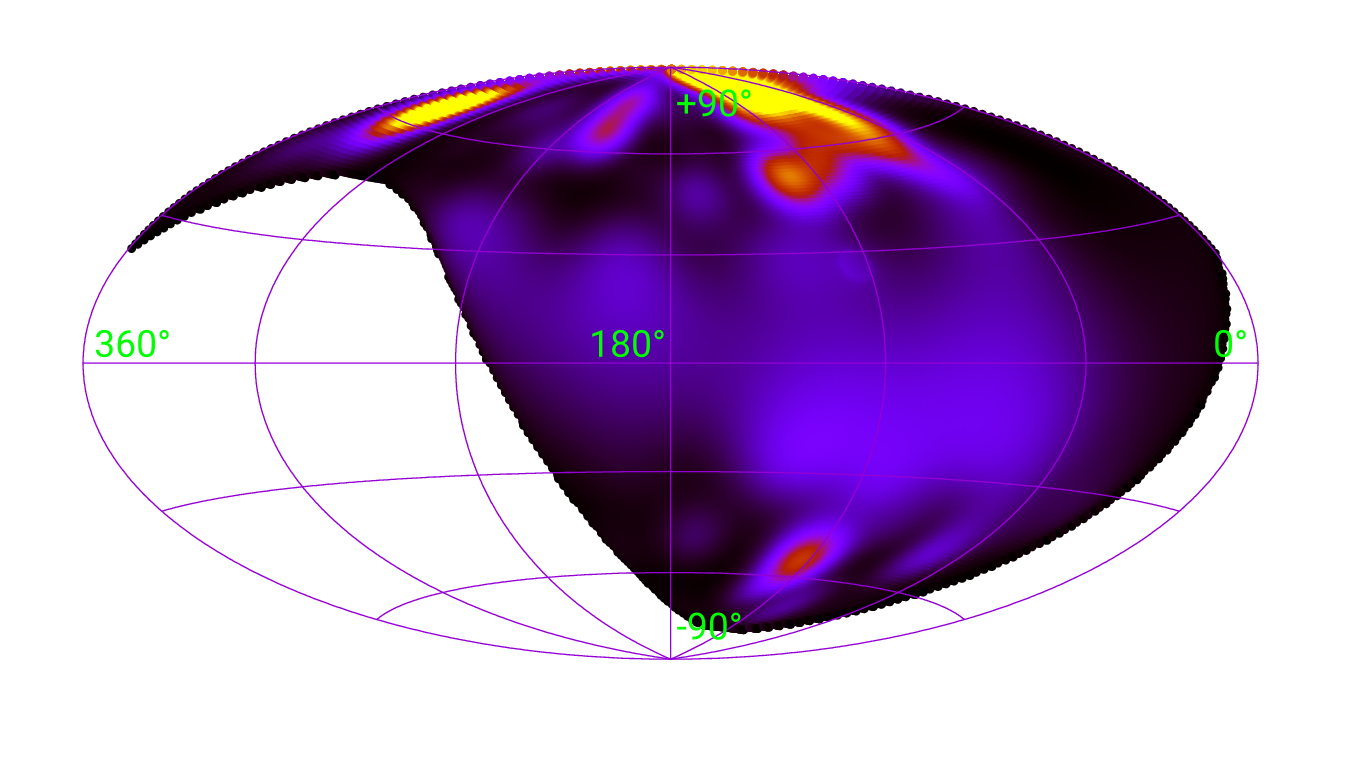}
\caption{
\label{fig:maps_p-Fe}
Examples of our basic UHECR flux model maps, $F_{{\rm p},k}$ and $F_{{\rm Fe},k}$ used for mock UHECR sets simulation. Deflections in regular GMF model of Ref.~\cite{Pshirkov:2011um} and b-dependent smearing in a random GMF model of Ref.~\cite{Pshirkov:2013wka} are assumed. No EGMF deflections.
{\it Left column:} protons with energies 10, 50, 100 and 200~EeV (from top to bottom).
{\it Right column:} iron with energies 10, 50, 100 and 200~EeV (from top to bottom).
Maps are shown in galactic coordinates for TA SD field of view.
}
\end{figure*}

We now discuss the details of the generation of the mock event sets that are used to compare a given model to observations. 
We first compute the flux maps for different injected species at different observation energies 
taking into account the UHECR injection spectrum, propagation, deflections by the magnetic fields and the detector effects. 
Thus we get a set of basic maps for various primaries at various energies.
Then we combine these maps with fractions of primaries corresponding to a particular composition model and use 
the resulting map to generate mock UHECR event sets.
The construction of the basic flux maps $F_{i, k}$, where $i$ denotes the injected particle type and $k$ the detected particle energy, is organized as follows.

UHECR sources are assumed to trace the luminous matter distribution in the Universe.
This can be achieved, on a statistical basis, by 
assigning each galaxy from a complete volume-limited sample an equal {\em intrinsic}
luminosity in UHECR. In practice, we use instead a flux-limited galaxy sample derived from the 2MRS galaxy catalog~\cite{Huchra:2011ii}. We cut out dim galaxies with ${\rm mag} >
12.5$ so as to obtain a flux-limited sample with a high degree of
completeness, and eliminate galaxies beyond $250$~Mpc. We assign progressively
larger flux to more distant galaxies to compensate for the observational selection
inherent in a flux-limited sample (see Ref.~\cite{Koers:2009pd} for the exact procedure).
In a similar way, we assign larger weights to the galaxies within $\pm
5^\circ$ from the Galactic plane to compensate for the catalog incompleteness in this region.
We also cut out galaxies at distances closer than $5$~Mpc as they are too few
to be treated statistically (this is equivalent to assuming that there are no sources
closer than this distance; if such sources exist they have to be added individually). 
Finally, we assume that sources beyond $250$~Mpc are distributed uniformly
with the same mean density as those within this distance.
The space distribution of sources obtained in that way is completely fixed.
The source number density, $\rho$, in this model is corresponding to that of all galaxies: $\rho \simeq 10^{-2}$~Mpc$^{-3}$~\cite{binney2008galactic}.
It should be emphasized that we use this source distribution for both
the generation of basic UHECR mock sets and for the computation of the TS, see next Subsection. 

The source number densities as low as $\rho \simeq 10^{-5}$~Mpc$^{-3}$ and even lower are not excluded experimentally~\cite{PierreAuger:2013waq} (see, however, recent studies that are placing more stringent limits~\cite{Kuznetsov:2023jfw, Bister:2023icg}). In case of such rare sources one would expect that the TS based on the catalog of {\it all galaxies} would show lower sensitivity to mass composition. In Sec.~\ref{sec:uncert_sources} we describe mock set simulations for low source number density and discuss this issue quantitatively.

We set UHECR injection spectra by fitting the TA and Auger observed spectra with the SimProp~v2r4~\cite{Aloisio:2017iyh} propagated spectra for each primary separately.
As a result, the following spectra are adopted for our basic expected UHECR flux: power law
with the indexes $-2.55$, $-2.20$, $-2.10$ and without injected energy cut-off for {\it protons},
{\it helium} and {\it oxygen}, respectively; power law with the index $-1.50$ and
with a sharp cut-off at $280$~EeV for {\it silicon}; power law with the index
$-1.95$ and with a sharp cut-off at $560$~EeV for {\it iron}.
The spectra for protons, helium and iron are derived from the fits
to TA observed spectrum~\cite{Ivanov:2020rqn}, while the spectra for oxygen and silicon
are adapted from Ref.~\cite{diMatteo:2017dtg}, where combined fits to TA~\cite{Tsunesada:2017aaq} and Auger~\cite{Fenu:2017hlc} were performed, taking into account an energy rescaling between the two experiments~\cite{Dembinski:2017zsh}.
Note that the shape of the injected spectrum cutoff is not important in our setup, according to discussion of Ref.~\cite{diMatteo:2017dtg}.
We show some examples and details of the spectra fitting in the Appendix.
The secondary protons generated
during propagation of injected primary nuclei through the interstellar medium
are taken into account for helium and oxygen nuclei. We adapt the method and the approximations
used in Ref.~\cite{diMatteo:2017dtg}. In particular, we assume that all nuclei
of atomic weight A injected with $E > 10 A$~EeV immediately disintegrate into $A$
protons having the energy $1/A$ times the injected energy of the nucleus each.
For a power-law injection of nuclei with index $\gamma$ and no cutoff,
this results in the following number of secondary protons $N_p$ above a given threshold $E_{\rm min}$:
\begin{equation}
N_p(\geq E_{\rm min}) = A^{2-\gamma} N_A(\geq E_{\rm min}).
\label{eq:secondaries}
\end{equation}
Because of the cutoff in the injection spectra the secondaries generated by silicon and iron nuclei drop out of the energy range  $E > 10$~EeV that we consider in this study.
More details on the approximations of UHECR propagation used are given in Ref.~\cite{diMatteo:2017dtg}.

We also found that for iron primary the observed spectrum can be fitted almost equally well by the injection with and without cutoff. For no-cutoff spectrum the injection slope is $-1.89$ and the observed flux is supplemented with secondary protons.
We choose the injection for iron with the cutoff --- this choice is conservative because it yields larger mean deflections.
In Sec.~\ref{sec:uncert_spec} we study how our results change if we use the no-cutoff iron injection instead.
We also discuss the effect of varying spectral indexes within their uncertainties.

Finally, following the approximations of Ref.~\cite{diMatteo:2017dtg} we assume the remnants of the primary nuclei, that are attenuated upon the propagation through the interstellar medium, at detection have the same charge as primary nuclei have at injection.
As it was shown in that study, this assumption lead to a per cent level errors with respect to full MC simulation of the propagation.
Moreover, in context of our study these corrections would act in a conservative direction, making the simulated deflections larger and the composition models more compatible with the data (see next Section).

We also consider an injection composition model from the Auger study~\cite{Aab:2016zth}.
Namely, we use their best-fit model with a power-law spectrum $E^{-0.96}$ and a rigidity dependent exponential cutoff of the special form.
The fractions of separate mass components are fixed at 1~EeV: $f_p = 0$, $f_{He} = 0.673$, $f_N = 0.281$, $f_{Si} = 0.046$, $f_{Fe} = 0$.
To get the appropriate spectrum at Earth taking into account the attenuation and secondaries, we use the results of the propagation for this model obtained with the code of Ref.~\cite{PierreAuger:2022axr}.
The results are obtained in the energy range $32 \leq E \leq 80$~EeV due to the limitations of the mentioned code.
The deflections in the GMF (see below) for this model are estimated according to an average charge of the observed composition at a given energy.

As we plan to use deflections of UHECRs from their sources as a variable discriminating between particle types, the effect of cosmic magnetic fields on the expected UHECR flux is of primary importance. 
The UHECR deflections by the galactic magnetic fields are implemented as follows.
In general, the galactic magnetic field has regular and random components.
For the regular field, we adapt the model of Ref.~\cite{Pshirkov:2011um} for our basic UHECR flux picture and the model of Ref.~\cite{Jansson:2012pc} for the test of result robustness.
The correction of the UHECR flux for the deflections in regular GMF is done by the standard backtracking technique.
The deflections in random magnetic fields, both Galactic and extragalactic, are modeled as smearing of the flux with the von Mises-Fischer distribution $f_\theta(\alpha)$ defined as 
\begin{equation}
f_\theta(\alpha) = \frac{\exp(2\cos\alpha/\theta^2)}{2\pi\theta^2\sinh (2/\theta^2)},
\label{eq:smearing}
\end{equation}
where the parameter $\theta$ is the smearing angle.
The magnitude of the smearing is proportional to the combination $Bq/E$ and is
different for UHECR species of different charges $q$ and energies $E$.

The galactic random field is non-uniform over the sky: the dependence of mean deflections $\sqrt{ \langle \theta^2 \rangle}$ (equivalently, the smearing angle) on the Galactic latitude has been
estimated from the dispersion of Faraday rotation measures of
extragalactic sources in Ref.~\cite{Pshirkov:2013wka}.
The following empiric relation has been obtained for protons of $E=40$~EeV:
\begin{equation}
\sqrt{ \langle \theta^2\rangle} \leq \frac{1^\circ}{\sin^2b +0.15},
\label{eq:rand_GMF}
\end{equation}
$b$ being the Galactic latitude.
Note that this formula is purely phenomenological and independent of any assumptions about morphology or coherence length of random GMF.
We adopt this relation conservatively treating it as the equality (i.e,
assuming maximum deflections) and rescaling it for other species and energies according to magnetic rigidity.
Subtleties of implementation of a non-uniform smearing are described in Ref.~\cite{Kuznetsov:2020hso}.

The deflections in extra-galactic magnetic field are set to zero in our basic flux model. This corresponds to either $B_{\rm EGMF} \ll 1$~nG for the correlation length $\lambda \sim 1$~Mpc or $B_{\rm EGMF} \ll 0.1$~nG for cosmological scale $\lambda$. The detailed discussion of possible UHECR deflections in EGMF, as well as quantitative estimate of their effect on our results, is given in Sec.~\ref{sec:uncert_EGMF}.

Finally, we add instrumental effects to our flux maps in order to fully reproduce the observed UHECR flux picture. We add a uniform smearing by $1^\circ$
to account for the angular resolution of TA. 
This only slightly affects flux maps for protons at high energies in the Galactic pole regions
where the deflections due to random and regular GMF may become comparable to $1^\circ$.
We should also note that the accuracy of our procedure of flux map construction is also $1^\circ$, that defines the overall accuracy of our method.
Finally, we modulate the flux maps by the geometrical exposure of the TA SD. We bin the energy in 20 logarithmic bins per decade starting from $5$~EeV, with the highest energy bin being an open interval $E > 200$~EeV, and generate a flux map for each injected species and each energy bin. 

Several examples of resulting model flux maps $F_{i,k}$ for injected protons and iron and for different energies are shown in Fig.~\ref{fig:maps_p-Fe}. 
Each map is a continuous function of the direction that is normalized to a unit integral over the sphere. It can be interpreted as a probability density to observe
an event from the direction ${\bf n}$. Given the flux map $F_{i,k}$ it is straightforward to generate the set of UHECR events that follow the corresponding distribution by throwing random events and accepting them with the probability $F_{i,k}({\bf n})$ according to their direction ${\bf n}$.
We generate the energies of the events in a mock set according to the reconstructed
TA spectrum~\cite{Ivanov:2015pqx} and additionally smear the energies with the Gaussian function of a width corresponding to the TA SD energy resolution of $18\%$
(for the Auger best-fit composition model we do not perform this smearing in order not to narrow down the available energy range, which is already not wide).
Each event is thrown using the flux map of the given species and energy of the bin it falls into. We generate a large number of events in each mock event set so as to make the statistical uncertainty of the corresponding TS negligible. 

\subsection{Test-statistics}
\label{sec:analysis:TS}
\begin{figure*}
 \includegraphics[width=0.99\columnwidth]{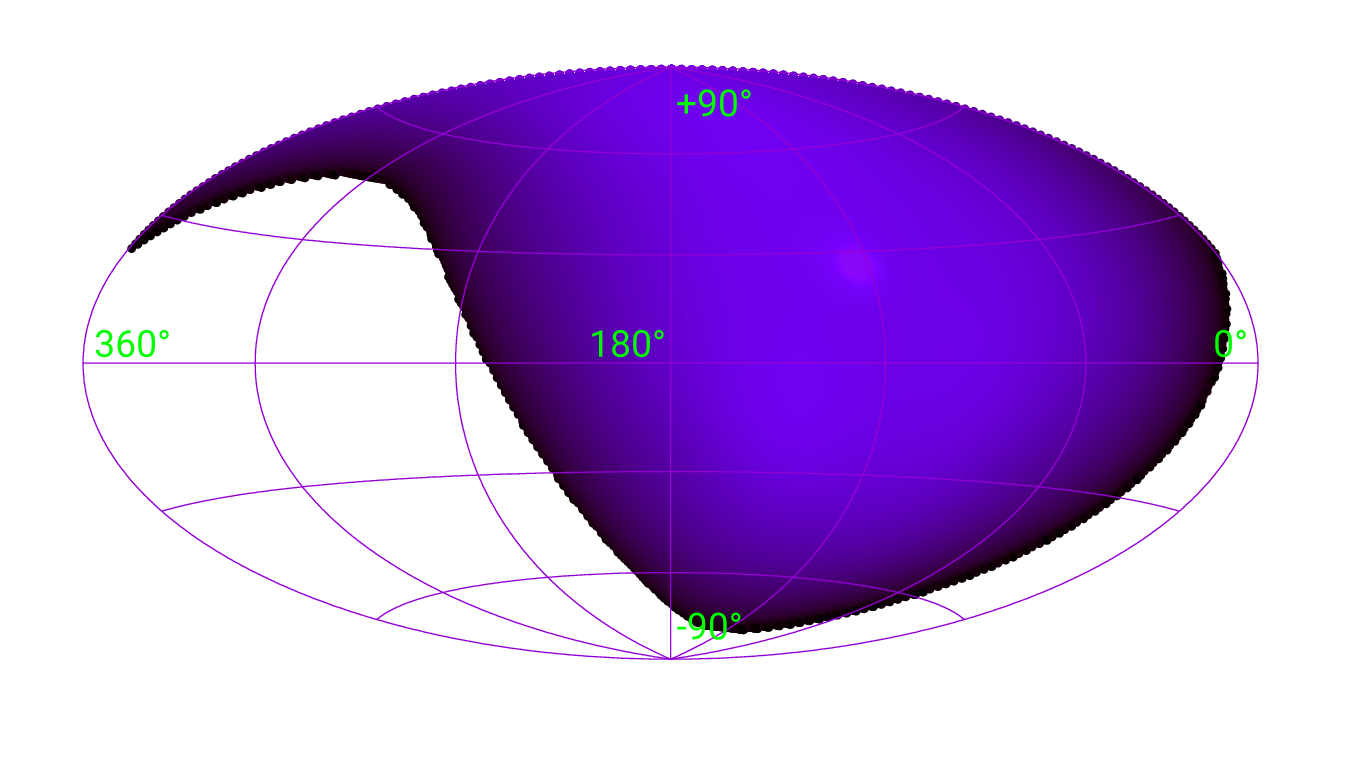}
 \includegraphics[width=0.99\columnwidth]{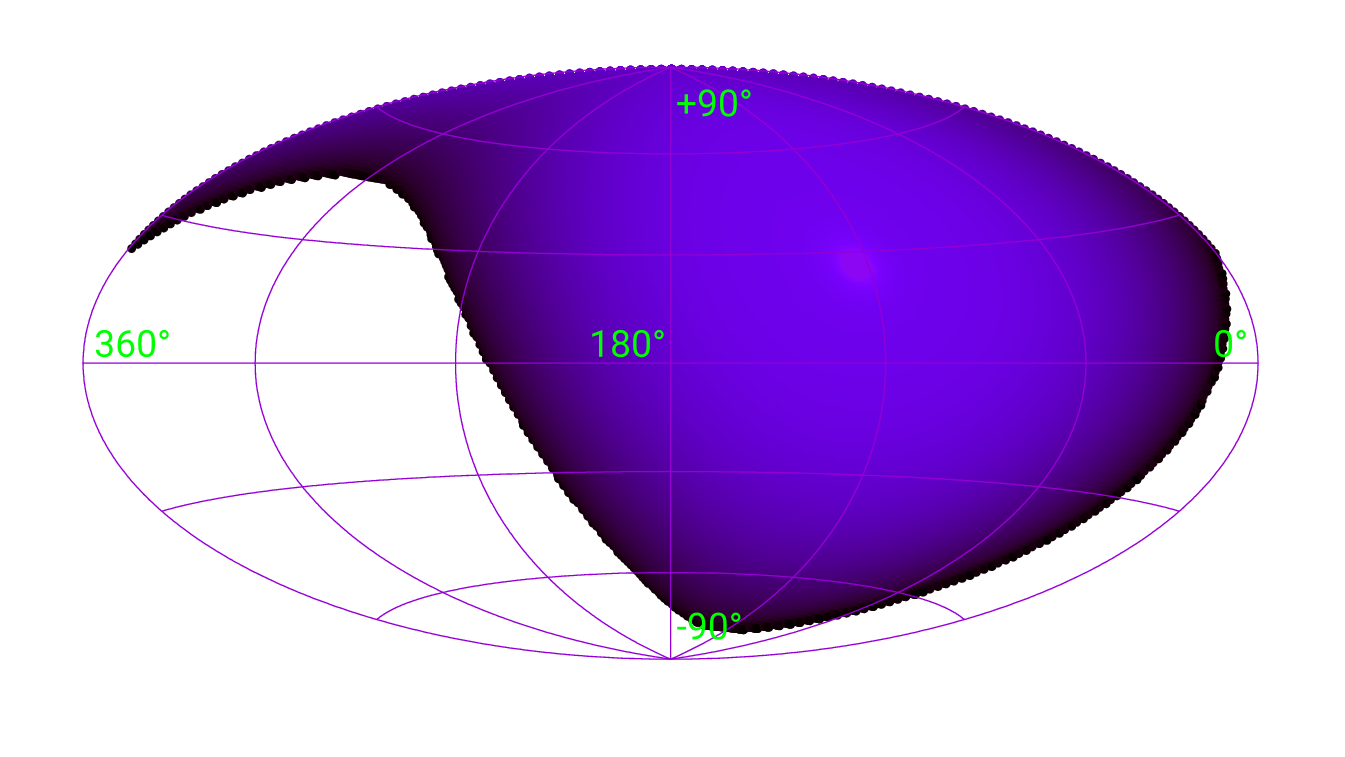}
 \includegraphics[width=0.99\columnwidth]{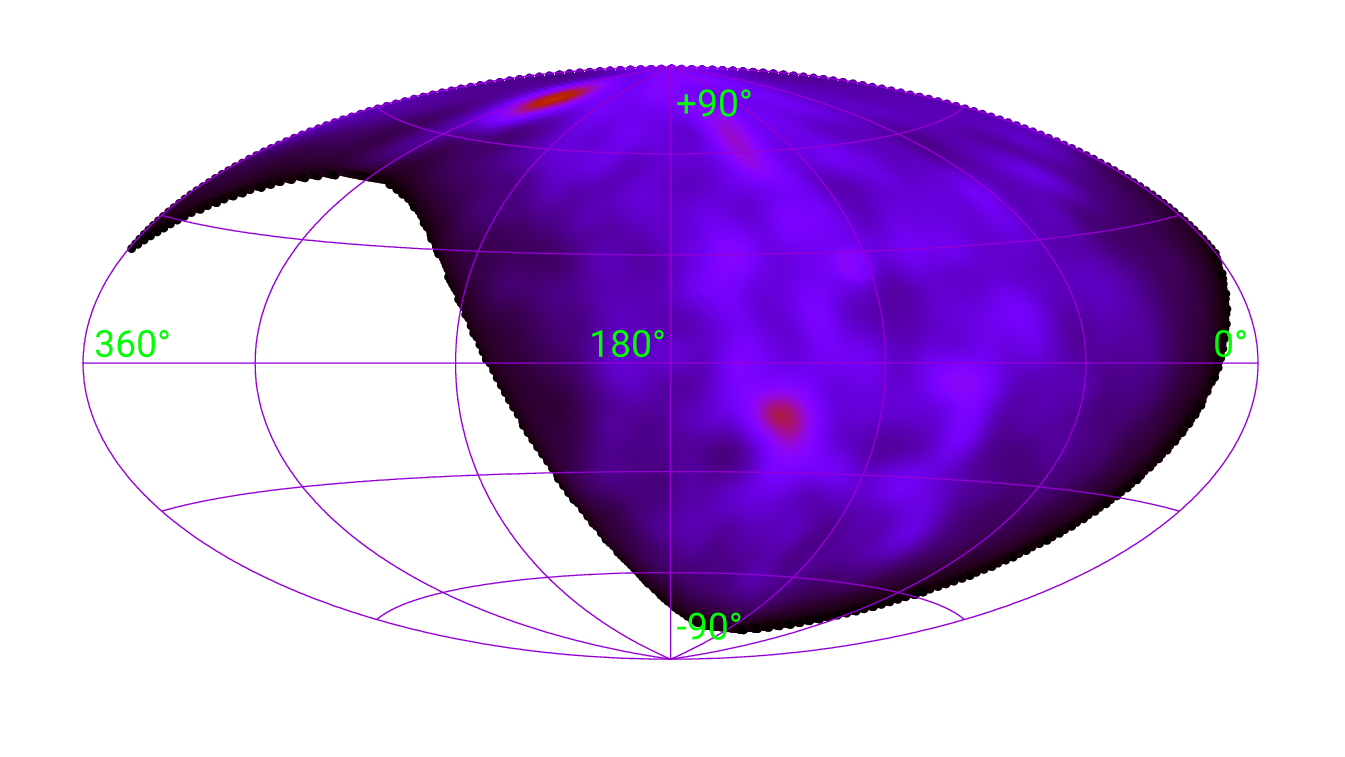}
 \includegraphics[width=0.99\columnwidth]{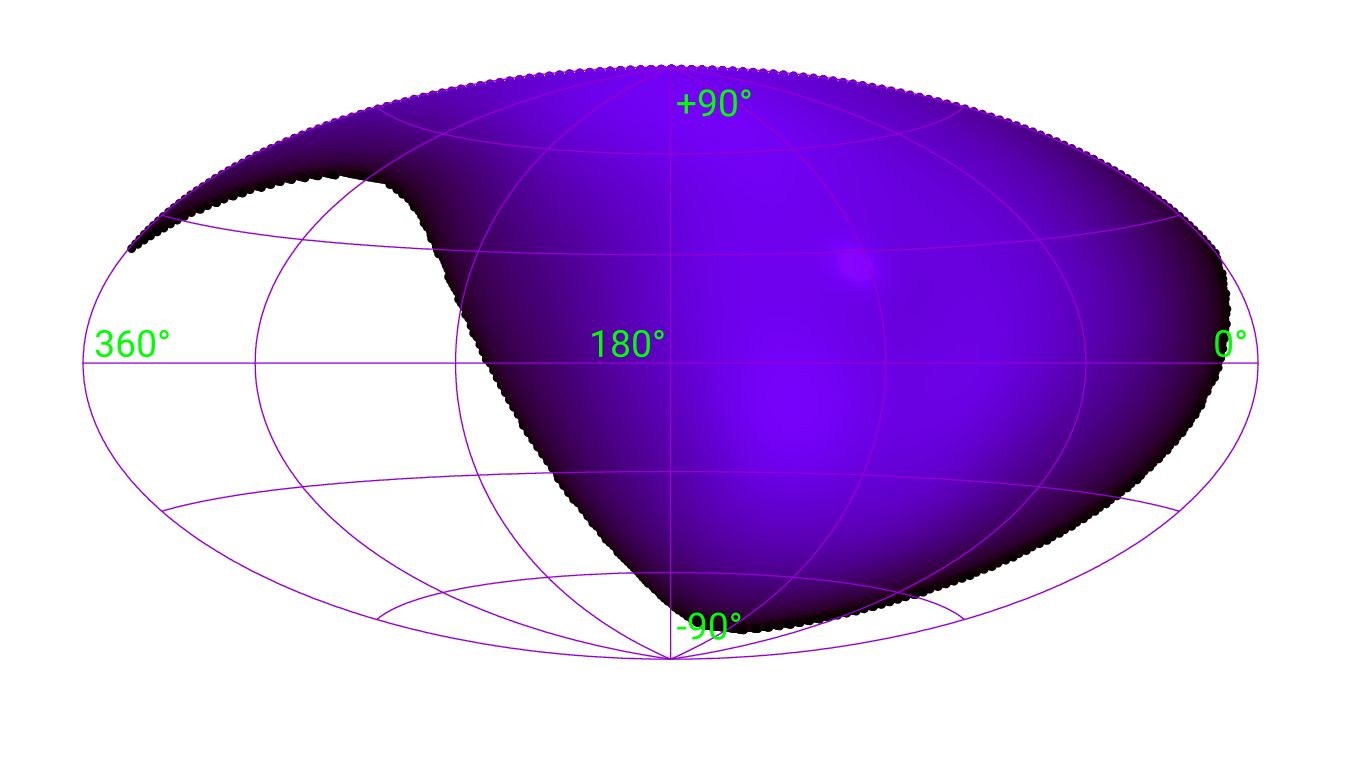}
 \includegraphics[width=0.99\columnwidth]{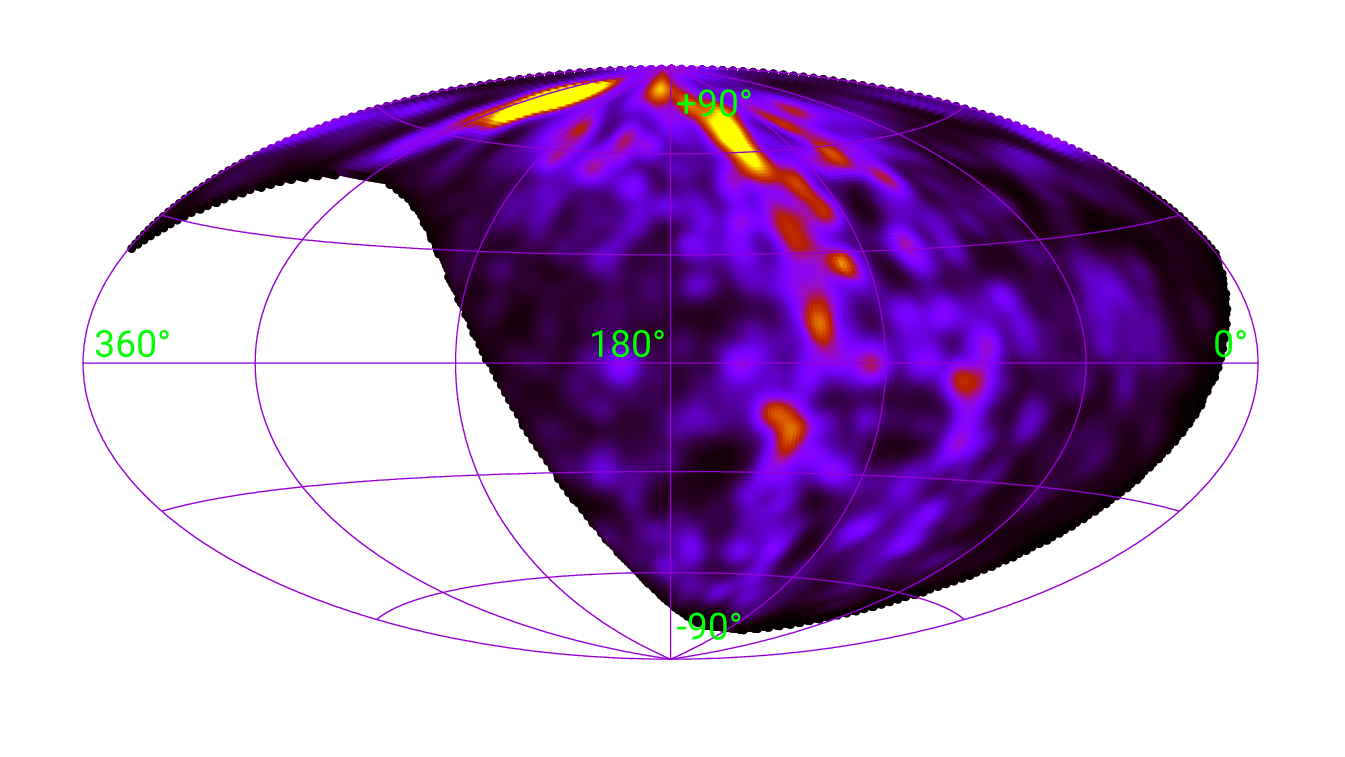}
 \includegraphics[width=0.99\columnwidth]{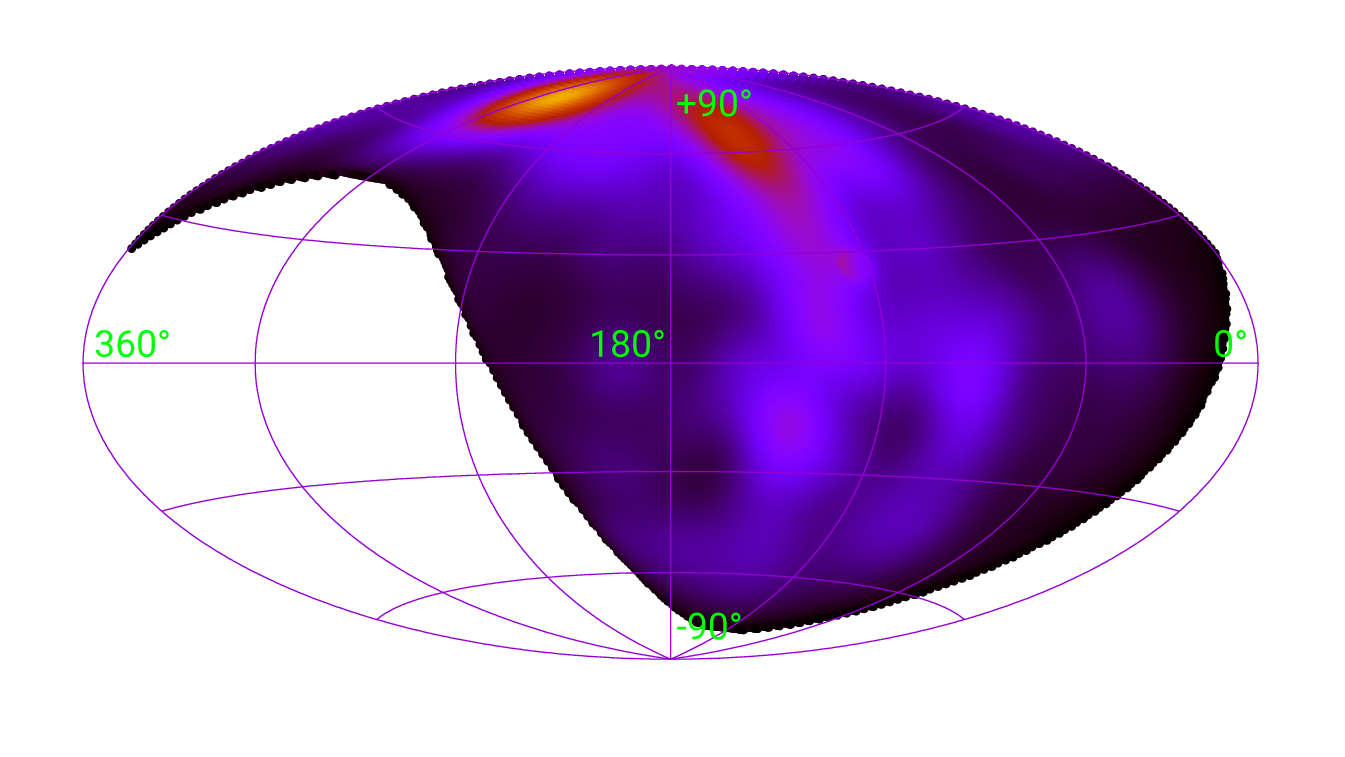}
 \includegraphics[width=0.99\columnwidth]{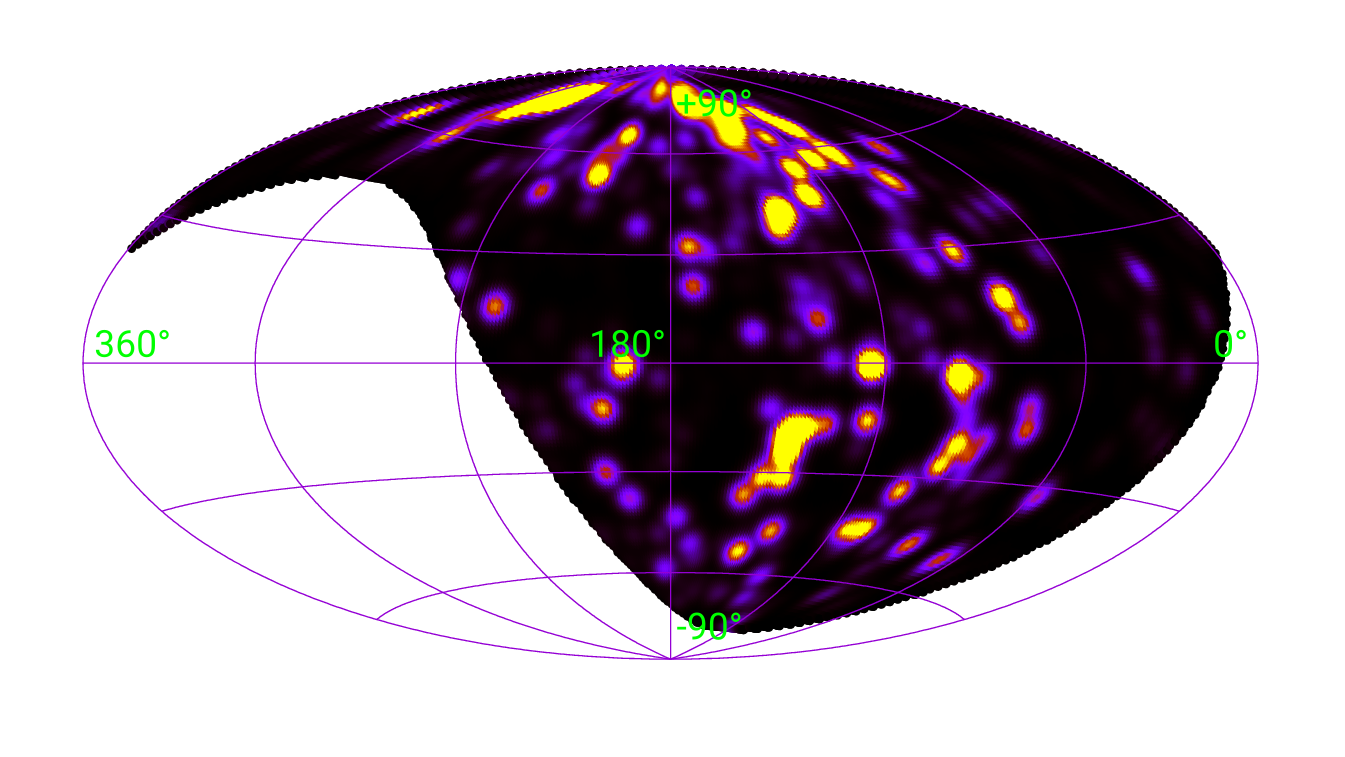}
 \includegraphics[width=0.99\columnwidth]{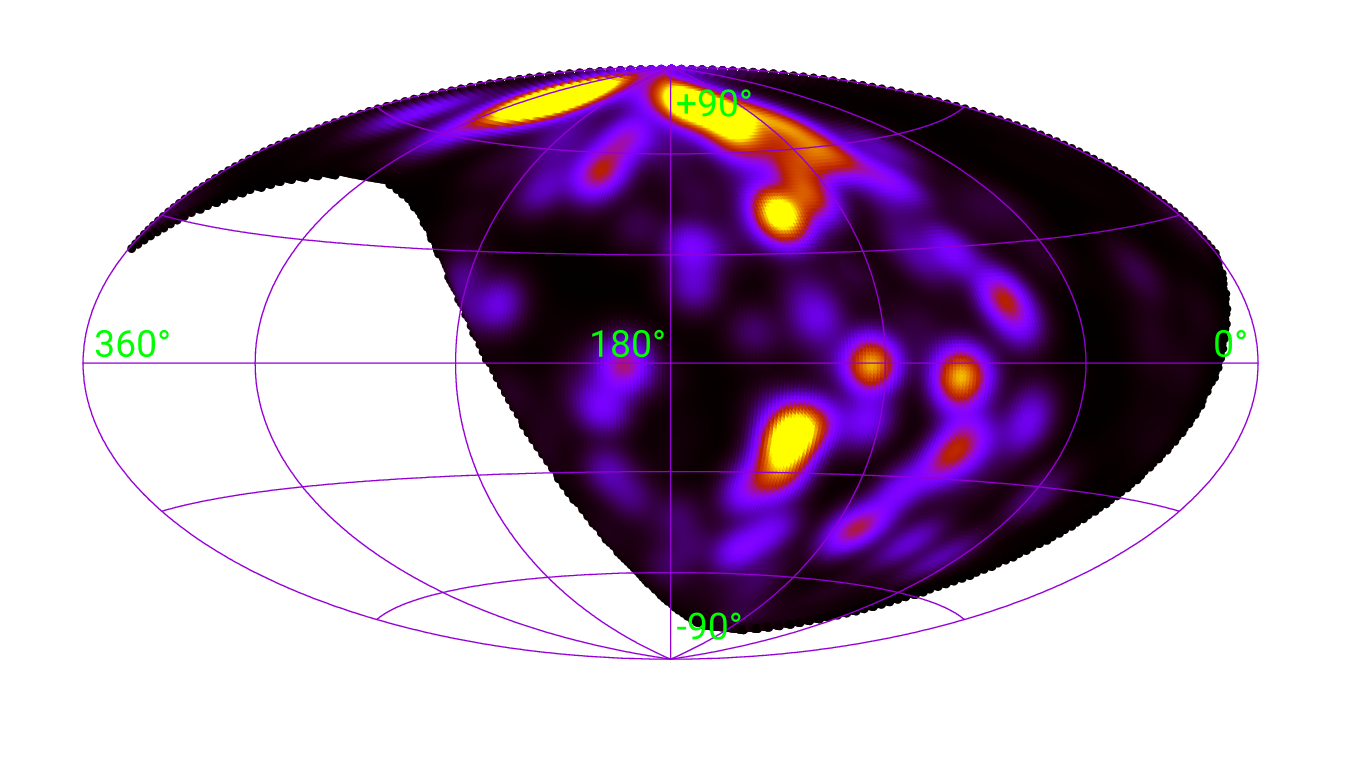}
\caption{
\label{fig:maps_TS}
Examples of flux maps for TS measurement, $\Phi_k$. Proton attenuation with injection spectrum index $\gamma = 2.55$ is assumed. Uniform smearing with magnitude $\theta_{100}$ at 100~EeV is applied.
{\it Left column:} $\theta_{100} = 2^\circ$ for energies 10, 50, 100 and 200~EeV (from top to bottom).
{\it Right column:} $\theta_{100} = 10^\circ$ for energies 10, 50, 100 and 200~EeV (from top to bottom).
Maps are shown in galactic coordinates for TA SD field of view.
}
\end{figure*}

The appropriate choice of the test statistics and the corresponding observable is very important for our method.
We want the TS to depend on the overall magnitude of deflections but be insensitive to their particular directions.
We would expect that such observable would not depend strongly on the details of the regular magnetic field, but mainly on its overall magnitude.
While the existing GMF models agree on the overall magnitude of the galactic field within $\sim 50$\%, the magnitude of deflections in various composition models differ from 1 to 26 times, according to particle charges.
Therefore we expect that the TS that is sensitive mainly to deflections magnitude would distinguish between different composition models despite the relatively poor knowledge of the galactic magnetic field.

Such an observable is inspired by the case of purely random UHECR deflections which are characterized by a single parameter, the width of the Gaussian spread of a point source.
More accurately, we use the von Mises-Fischer distribution~(\ref{eq:smearing}).
By analogy, we choose to characterize the given set of events by their typical deflection angle with respect to the sources in the LSS. To compute this quantity we construct another set of sky maps $\Phi_k(\theta_{100})$ that are simplified analogs of the flux model maps $F_{i, k}$. Namely, each map $\Phi_k(\theta_{100})$ is derived from the same LSS source distribution with the flux attenuated as protons with injection spectrum index $2.55$ taken at detected energy $E_k$ and uniformly smeared with the angle $\theta = (100~{\rm EeV}/E_k) \cdot \theta_{100}$, where $\theta_{100}$ is the composition-discriminating parameter to be determined from fitting to the data or mock sets: 
given the set of events with directions ${\bf n}_i$ one can determine the value of $\theta_{100}$ by computing the $\theta_{100}$-dependent test statistics 
\begin{equation}
TS(\theta_{100}) = -2 \sum_k \left( 
\sum_i \ln  \frac{ \Phi_k(\theta_{100}, {\bf n}_i)}
{\Phi_{\rm iso} ({\bf n}_i)}
\right). 
\label{eq:TS}
\end{equation}
Here the internal sum runs over the events in the energy bin $k$ and we have
included a standard normalization factor $-2$. For convenience we also
included the normalization factor $\Phi_{\rm iso}({\bf n}_i) = \Phi(\infty,{\bf n}_i)$
that corresponds to the isotropic distribution of sources --- a uniform flux map
modulated by the exposure function. The energy binning here is the same as for the model flux maps $F_{i, k}$.
The parameter $\theta_{100}$ ranges from $1^\circ$ to $200^\circ$, where the first value comes from
the experiment resolution and the second one corresponds to the size of the TA field of view (FoV) and 
mimics the isotropic distribution. One can infer the value of $\theta_{100}$
for the given event set by finding the TS minimum with respect to it.  This minimum,
$\theta_{100}^{\rm min}$, is interpreted as the typical deflection angle with respect
to the sources in the LSS. The width of the minimum, $\sigma(TS(\theta_{100}))$ characterizes
the uncertainty of the deflection angle, and the square root of the minimum depth,
$|TS(\theta_{100}^{\rm min})|^{1/2}$, measures the significance of the departure
of a given set from isotropy in standard deviations. A detailed discussion
of the TS choice and construction is given in the study~\cite{Kuznetsov:2020hso}.
Several examples of maps $\Phi_k(\theta_{100})$ are shown in Fig.~\ref{fig:maps_TS}.

We should stress that one and the same TS is used to quantify any mock event set with arbitrary injected composition, or a data set. The applicability of such a TS for event sets generated with different assumptions about UHECR flux is justified by the tests of the statistical power of the TS in distinguishing these event sets. These tests were performed in Ref.~\cite{Kuznetsov:2020hso} as well as in the present study (see next two sections).

For a sufficiently large event set the TS yields a deep and narrow minimum
at some value $\theta_{100}^{\rm min}$ that, within our approach, is a single characteristics
of a given composition model. Comparing the values of $\theta_{100}^{\rm min}$ for various models
with the TS distribution for the data one may determine to what extend each of these models
is compatible with the data. To make the picture more precise we estimate the compatibility
separately is several energy ranges (not to be confused with the technical energy bins $E_k$ used for TS construction).
Namely, we use the logarithmic energy ranges of 0.25 decade starting from $10$~EeV,
with the fifth range being the open interval $E > 100$~EeV.

\section{Results}
\label{sec:results}
\begin{figure}
 \includegraphics[width=0.99\columnwidth]{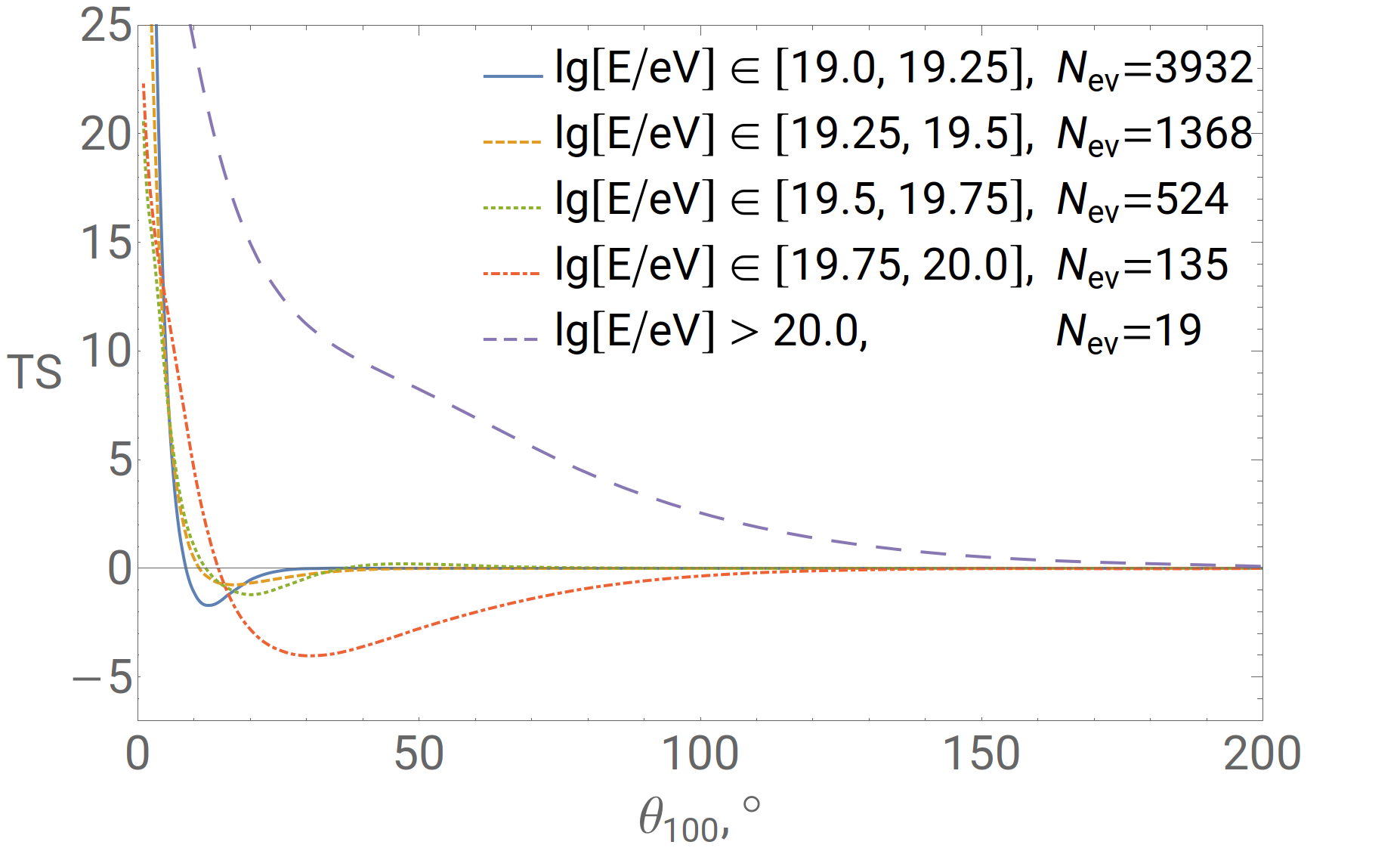}
\caption{
\label{fig:TS-data}
The distribution of test-statistics over $\theta_{100}$
evaluated for TA SD experimental data in five energy bins. The number of events in each bin is shown, in the legend.}
\end{figure}

\begin{figure*}
 \includegraphics[width=0.99\columnwidth]{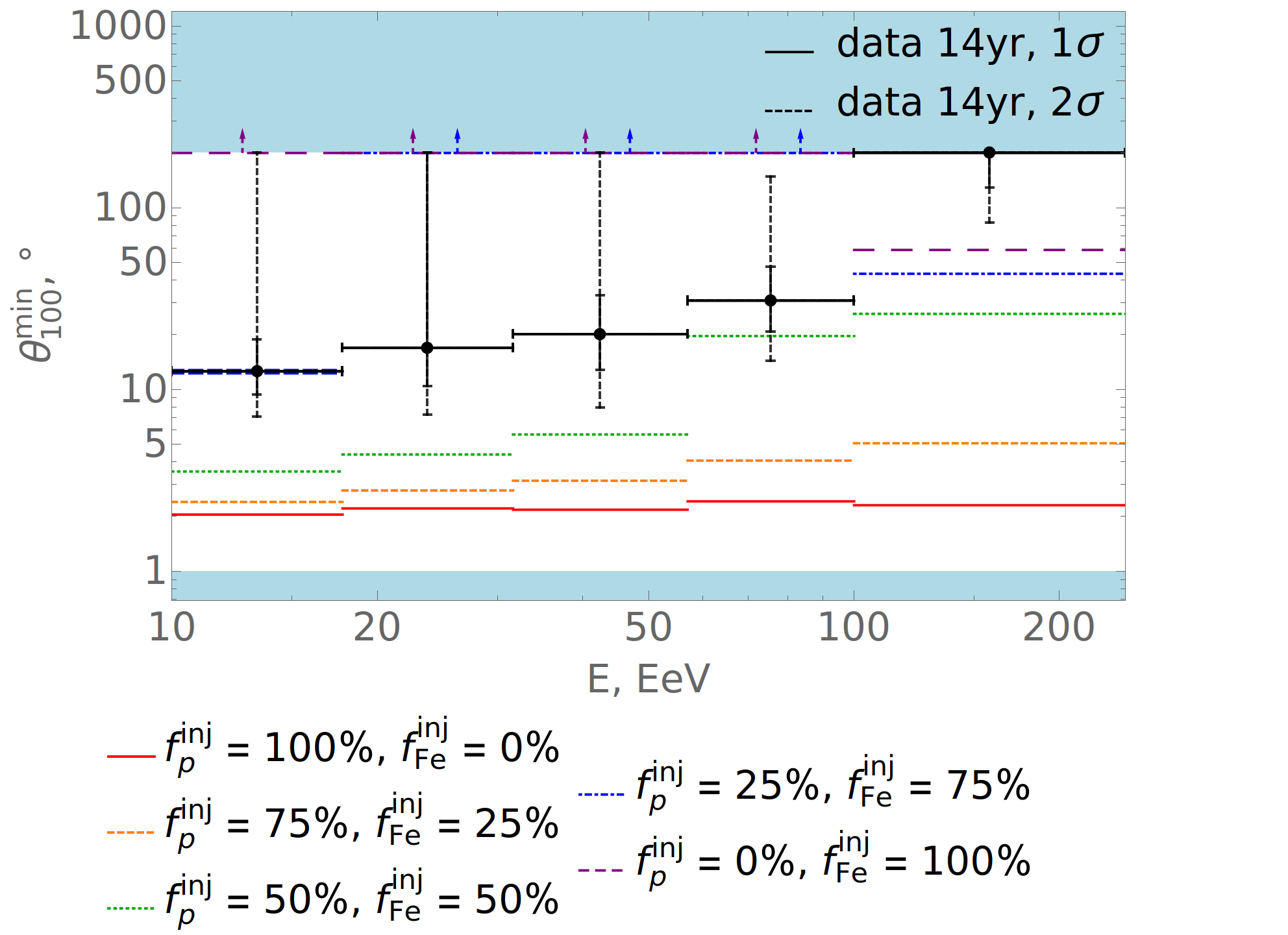}
 \includegraphics[width=0.99\columnwidth]{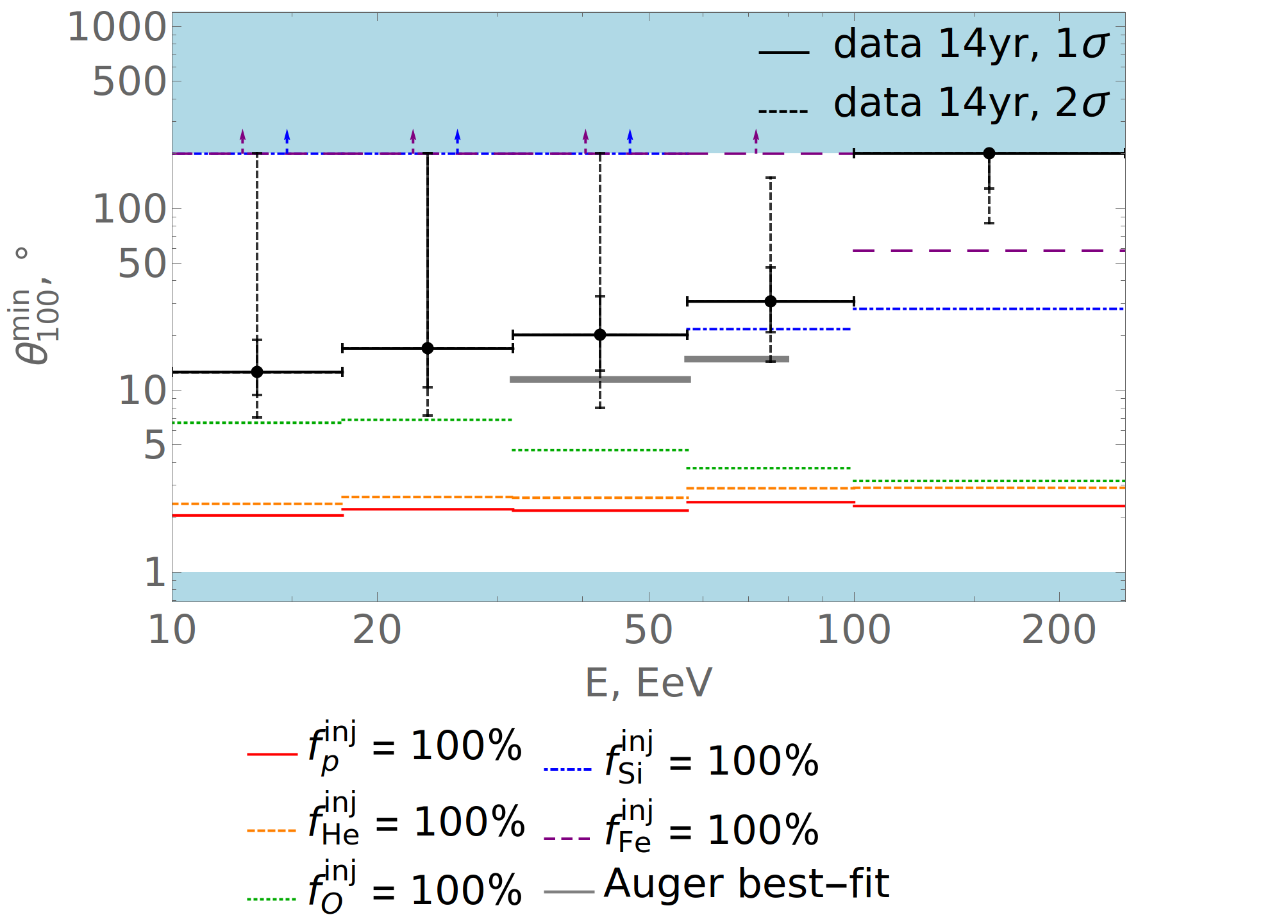}
\caption{
\label{fig:rails}
The test-statistics for the data compared with several injected composition models.
Regular GMF model of Ref.~\cite{Pshirkov:2011um} and random GMF model of Ref.~\cite{Pshirkov:2013wka} is used and deflections in EGMF are neglected.
Note that several heavy composition models yield the same value of 
$\theta_{100}^{\rm min} = 200^\circ$, i.e. they are indistinguishable in our method.
The corresponding lines that merge together on the plot are indicated by arrows.
{\it Left panel:} p-Fe mix composition models. {\it Right panel:} pure nuclei composition models and Auger best-fit composition model of Ref.~\cite{Aab:2016zth} (see text).}
\end{figure*}

The distributions of the TS for the TA SD data in five energy ranges is shown in Fig.~\ref{fig:TS-data}. We stress again that {\em the same} parameter $\theta_{100}$ --- the typical deflection rescaled to the energy $E=100$~EeV --- is measured in different energy ranges. 
Even before comparing these distributions with simulated models one can notice three important points. First, all the curves show steep rise at small values of $\theta_{100}$. This implies that at all considered energies the data is incompatible with small deflections of events from the LSS. Second, while at low energies the data does not show any clear preference for any deflection magnitude --- all the minima are shallow, if present at all --- at $56 \lesssim E \leq 100$~EeV the minimum exists at $\theta_{100}^{\rm min} = 30.8 ^\circ$, implying that data exhibit the correlation with the LSS at more than $2 \sigma$ level as compared to isotropy. Note that this value is global and should not be penalized, as no scanning in any parameters was performed.
Finally, at energies $E > 100$~EeV the data shows no hint of a minimum and prefers complete isotropy. This last remarkable feature is discussed and physically interpreted in our companion letter~\cite{TelescopeArray:2024oux}.

In Fig.~\ref{fig:rails} we confront the same data TS distributions with the values of $\theta_{100}^{\rm min}$ of various pure and mixed composition models.
As the TS is non-Gaussian we explicitly show the $1 \sigma$ and $2 \sigma$ error-bars around the TS minima that are shown with black points. We adjust the statistics of mock event sets 
to be $\sim 1000$ times larger than that of the real data in each
energy bin so that statistical uncertainties of the model predictions are negligible.

One can see that, 
with our basic assumptions about the UHECR flux model, the light or even the intermediate mass composition
is in tension with data. The situation is even more interesting at highest energies, where even pure iron is hardly compatible with data.

Something else one can see from Fig.~\ref{fig:rails} is that the sensitivity of the proposed TS to composition models
is not constant in energy and is a competition of two different trends: the evolution of the expected flux with energy and the simultaneous change in events statistics.
At lower energies, the expected flux from the LSS is almost uniform with a very small density contrast
(modulo the experiment's exposure) due to large contribution of remote
uniformly distributed sources and larger
deflections in magnetic fields. At higher energies, on the contrary, the map contrast
increases greatly due to simultaneous shrinking of the UHECR horizon and
decrease of magnetic deflections. On the other hand, the statistics decreases at high energies. It appears that 
the first trend wins: even the
small event statistics at highest energies gives a better sensitivity in
terms of the mass composition discrimination than the large
event statistics at lower energies. 

The non-monotonic behavior of the model predictions from bin to bin is a result of a complicated interplay between
various factors affecting the evolution of the model flux maps with energy, such as the 
fraction of the isotropic component, the flux focusing and
secondary source images that might appear due to large deflections in the lower energy
flux maps, the ratio between the mean total deflection of a given mass component
at a given energy and the size of the TA FoV, etc. These effects make it difficult to
predict qualitatively the evolution of $\theta_{100}^{\rm min}$ with energy and
composition, especially in the case of more than one mass
component. Still, a global trend of $\theta_{100}^{\rm min}$ to increase with the 
mass of the injected particles is visible in each energy range. We should also mention that the results in separate bins of the {\it observed} energy are not completely independent, as they are projected to a partially overlapping bins of the {\it injected} energy.

It is also
visible that the TS has better model separation power for event sets
where the deflections of separate mass components do not differ
significantly. This is the main reason for the counter-intuitive result of the method's
higher model separation power at lower energies where all deflections are
higher: both proton and iron deflections are close to isotropic at low energies,
while at higher energies protons deflections are small but iron deflections
are still close to isotropic. The method reaches its best sensitivity at
highest energies $E > 100$~EeV, where the total mean deflections of all
studied composition components are within FoV and sources are more distinct in
the sky-map. Therefore, all the composition lines are below our adopted 
``isotropic'' value $\theta_{100}^{\rm min} = 200^\circ$.
In this case the lines are not degenerate, which allows us to distinguish between several strongly deflected composition models that are indistinguishable at lower energies.

\section{Uncertainties}
\label{sec:uncert}
In this section we discuss the impact of our theoretical assumptions
and experimental uncertainties on the mass composition constraints. Note that in 
our approach all the uncertainties have an impact on the positions of the model lines but do not affect the data points.
To estimate the impact of each uncertainty we 
compare the model predictions computed for the basic value of a given parameter with the predictions 
computed for a varied value of this parameter. The main sources of uncertainties are discussed below one by one.

\subsection{Injection spectra}
\label{sec:uncert_spec}
\begin{figure*}
 \includegraphics[width=0.99\columnwidth]{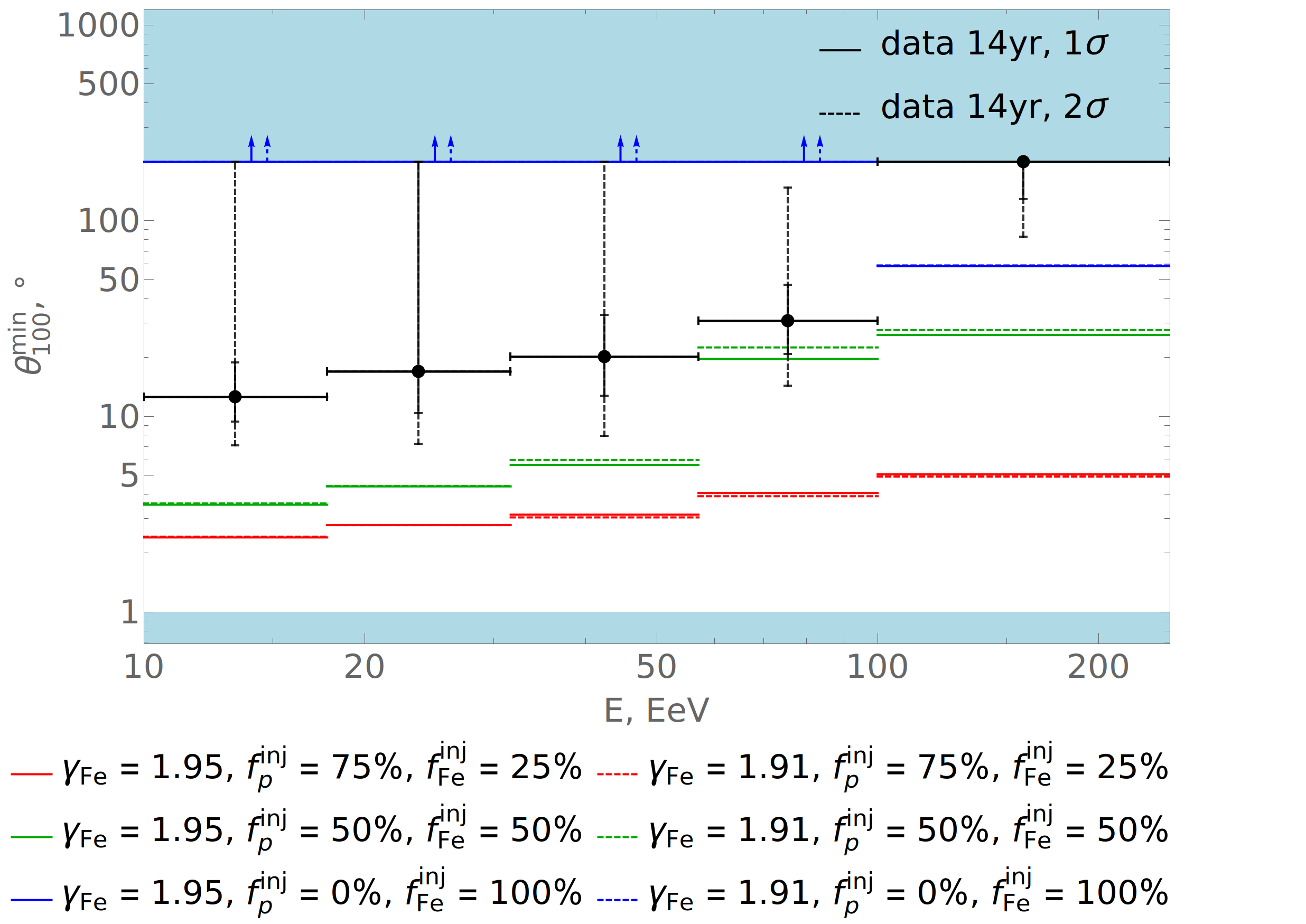}
 \includegraphics[width=0.99\columnwidth]{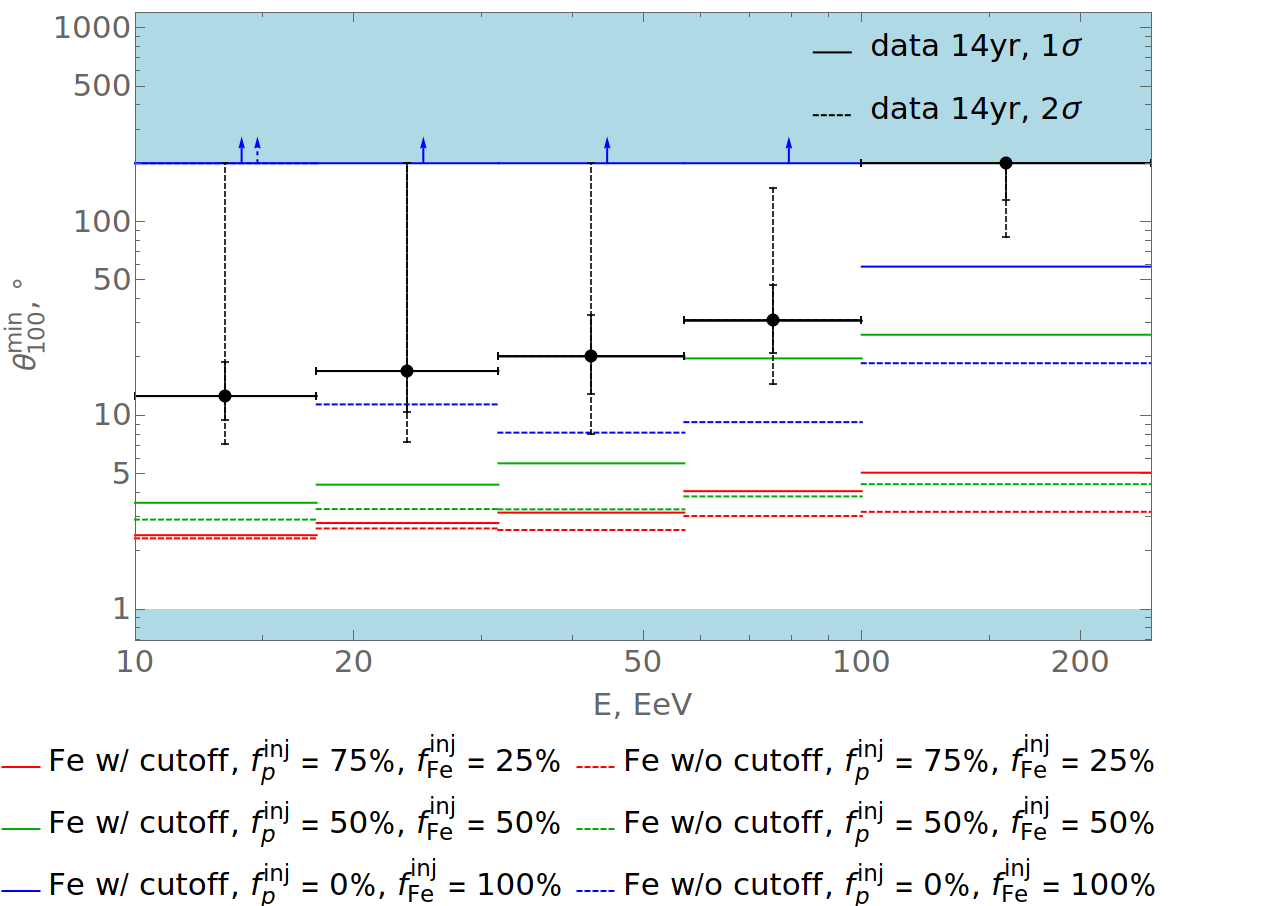}
\caption{
The TS for the data compared to some reference injected
composition models.
{\it Left panel:} p-Fe mix composition models for basic injection index values $\gamma_p = 2.55$, $\gamma_{Fe} = 1.95$ and for varied value $\gamma_p = 2.55$, $\gamma_{Fe} = 1.91$.
{\it Right panel:} p-Fe mix composition models for iron injection with cutoff at 560~EeV and without cutoff.
}
\label{fig:uncert_spec}
\end{figure*}
The injection spectra fits described in the Section~\ref{sec:analysis:sim}
yield a value of the spectrum index for each primary and a $1\sigma$ interval around it.
We compute the composition model lines varying the index within these intervals and compare the result with the basic lines.
We use proton-iron mix models for the tests of the uncertainties related to spectrum and energy scale.
The fitted values of injection spectrum indices are $2.55^{+0.04}_{-0.03}$ for protons and $1.95^{+0.04}_{-0.04}$ for iron.
We set the index for one primary to its best fit value and vary the index for another primary.
Among all the resulting models we choose the one with maximum deviation from the best-fit injection model. This happens to 
be the model with $\gamma_p = 2.55$ and $\gamma_{Fe} = 1.91$.
The resulting comparison is shown in Fig.~\ref{fig:uncert_spec}, left panel.
One can see that the impact of the variatio of the injection index on the model line position is negligible.

There is also an uncertainty associated to the presence of the cutoff in the injection spectrum for heavy primaries.
For heavy primaries the spectrum fits with and without cutoff are equally viable,
while leading to quite different expected flux model maps.
Therefore it is instructive to test the change in the composition results due to assumption of no-cutoff injection for iron.
The comparison is shown in Fig.~\ref{fig:uncert_spec}, right panel. One can see that when there is no cutoff in the iron spectrum, 
the predicted value of $\theta_{100}^{\rm min}$ is much lower (obviously, due to large fraction of secondary protons in the flux), so that for instance at $32 \lesssim E \lesssim 56$~EeV it is even hard to reconcile {\it any} composition with the data. As the data in general disfavors small deflections,  our basic model of iron injection with cutoff
(and hence without secondaries) is conservative.

\subsection{Systematic uncertainty of energy scale}
\label{sec:uncert_E}
\begin{figure*}
 \includegraphics[width=0.99\columnwidth]{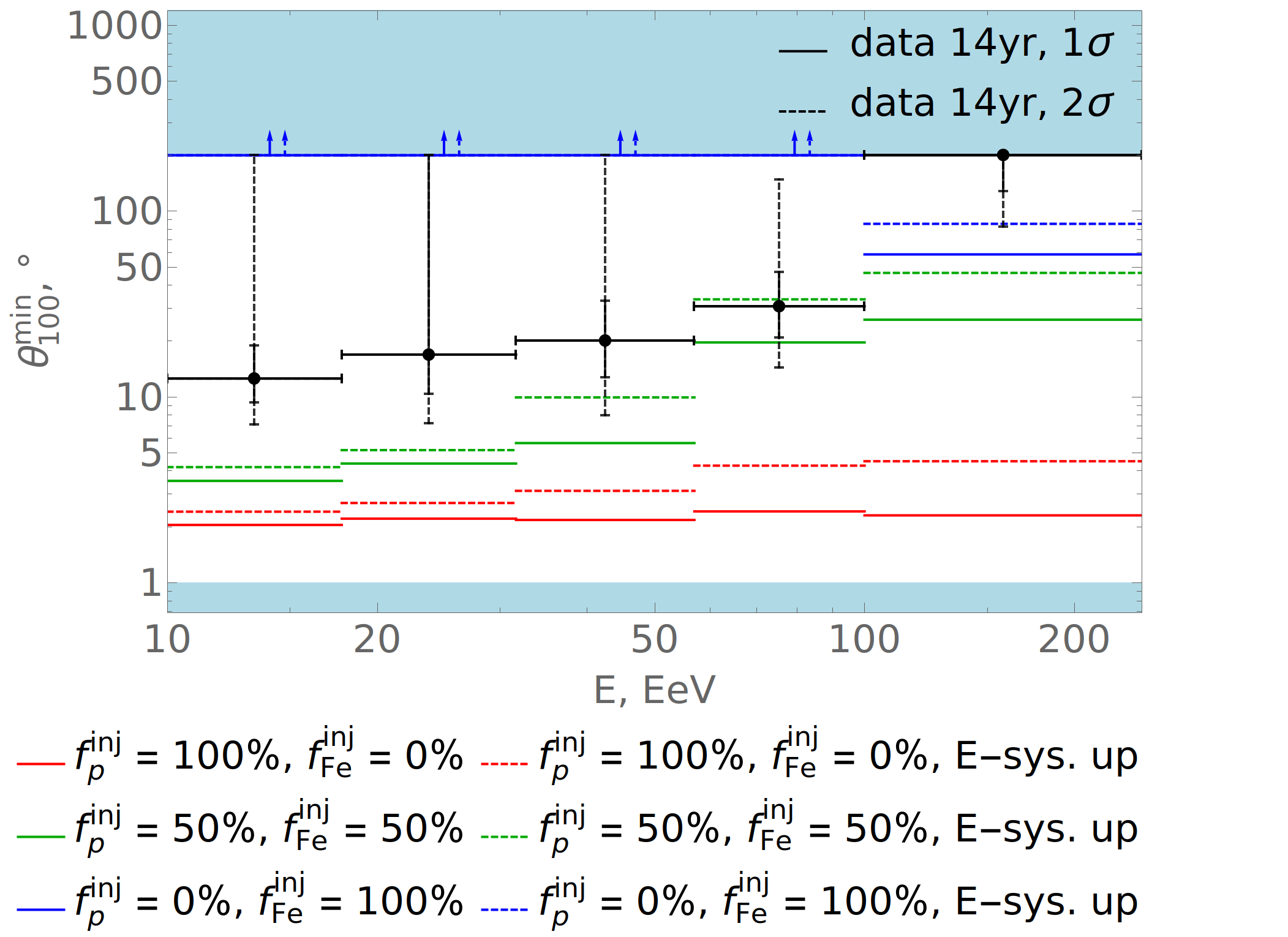}
 \includegraphics[width=0.99\columnwidth]{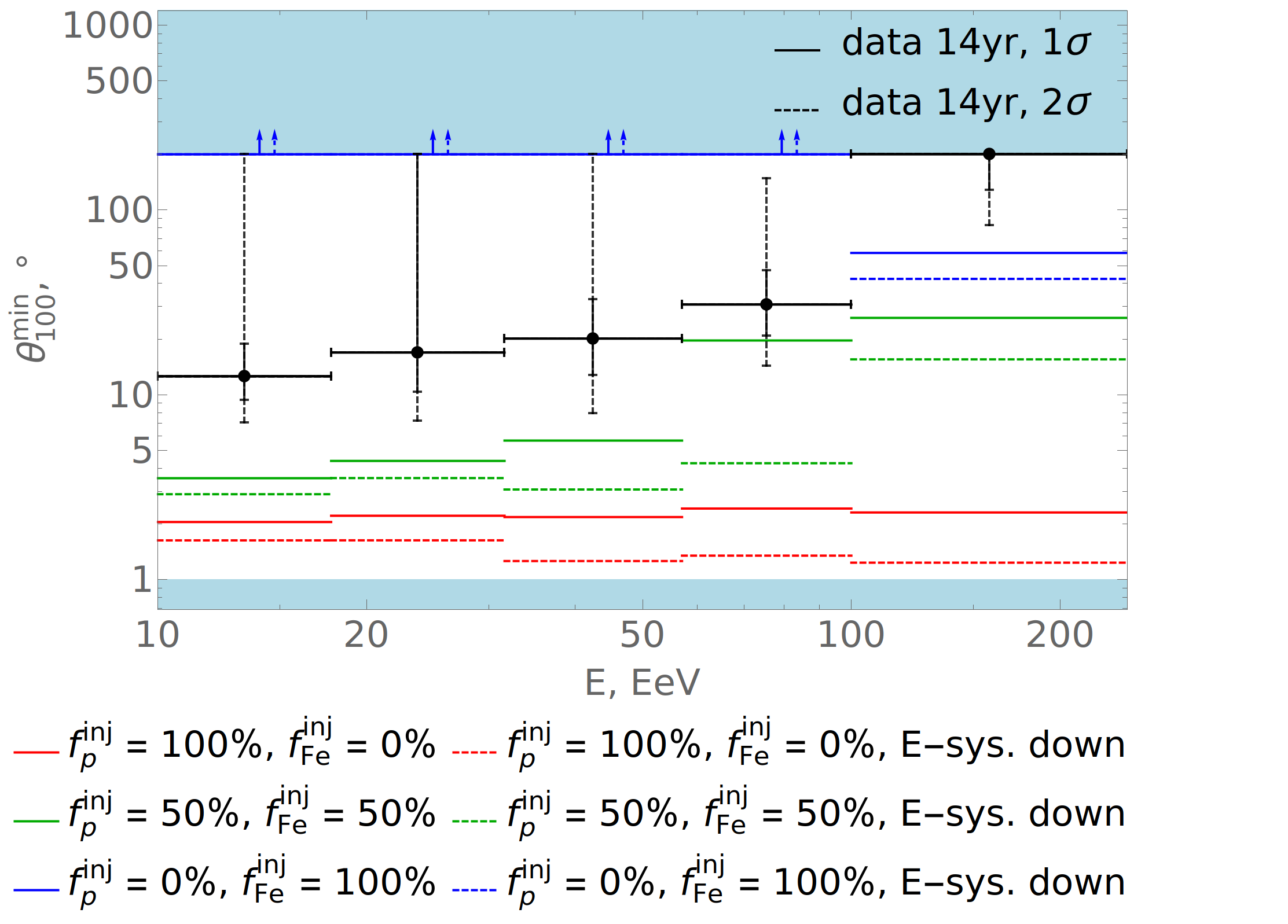}
\caption{
The test statistics for the data compared to some reference injected
composition models.
{\it Left panel:} p-Fe mix composition models with basic energy scale and with the energy scale shifted to the upper edge of its systematic band.
{\it Right panel:} the same but the energy scale is shifted to the lower edge of its systematic band.
}
\label{fig:uncert_E}
\end{figure*}

In the standard TA SD energy reconstruction procedure the overall energy scale is set to that established in the fluorescence measurements~\cite{AbuZayyad:2012ru}.
Therefore, a systematic uncertainty of the SD energy measurement is given by that of the FD energy scale which was found to be $21\%$~\cite{TheTelescopeArray:2015mgw}.
We estimate the impact of this uncertainty on our composition results by shifting the energies of all the events in a mock set to the lower or the upper edge of the systematic uncertainty band.
The results are shown in Fig.~\ref{fig:uncert_E}, where the left panel corresponds to the situation when measured energy is systematically higher than the real one and the right panel --- to the opposite situation.
One can see that the difference in model lines due to this uncertainty grows with energy, but does not exceed the difference between the light and heavy composition models.
It is also worth noting that the inconsistency of the data at $E > 100$~EeV with a light or intermediate composition is robust to this uncertainty.

\subsection{Galactic magnetic fields}
\label{sec:uncert_GMF}
A strength of the regular GMF component is known to be several $\mu G$ from Faraday rotation measures of extragalactic sources and from some other observations~\cite{Haverkorn2015}.
However, its general structure is unknown since a reconstruction of a 3D field from its 2D
projection on the sky is ambiguous.
Several proposed phenomenological models~\cite{Han:2006ci, Sun:2007mx, Pshirkov:2011um, Jansson:2012pc, 2017A&A...600A..29T} should be used with caution.
We also should note that some models predict quite large magnetic fields in the galactic halo~\cite{Shaw:2022lqd}. The estimated UHECR deflections in these fields for some directions in the sky can be enhanced significantly with respect to ``basic'' phenomenological models. However, these deflections are in general less than those expected in models of strong EGMF (see discussion in the next Subsection).

Our main initial motivation for a new TS (\ref{eq:TS}) was to minimize the impact of GMF uncertainty on the results of the composition estimation. It is therefore interesting to see to which extent this works in practice.
To estimate the impact of the GMF uncertainty we compare the TS predictions in one and the same composition model generated with our reference GMF model~\cite{Pshirkov:2011um} (PT'11) and with the model of Ref.~\cite{Jansson:2012pc} (JF'12).
The comparison is shown in Fig.~\ref{fig:uncert-MF}, left panel.
One can see that, as expected, the change in the predicted value of the TS with the change of the GMF model is small. 
Remarkably, in the majority of cases the predictions for the two GMF models are remaining compatible with the data within the same number of sigmas.

\subsection{Extra-galactic magnetic fields}
\label{sec:uncert_EGMF}
\begin{figure*}
 \includegraphics[width=0.99\columnwidth]{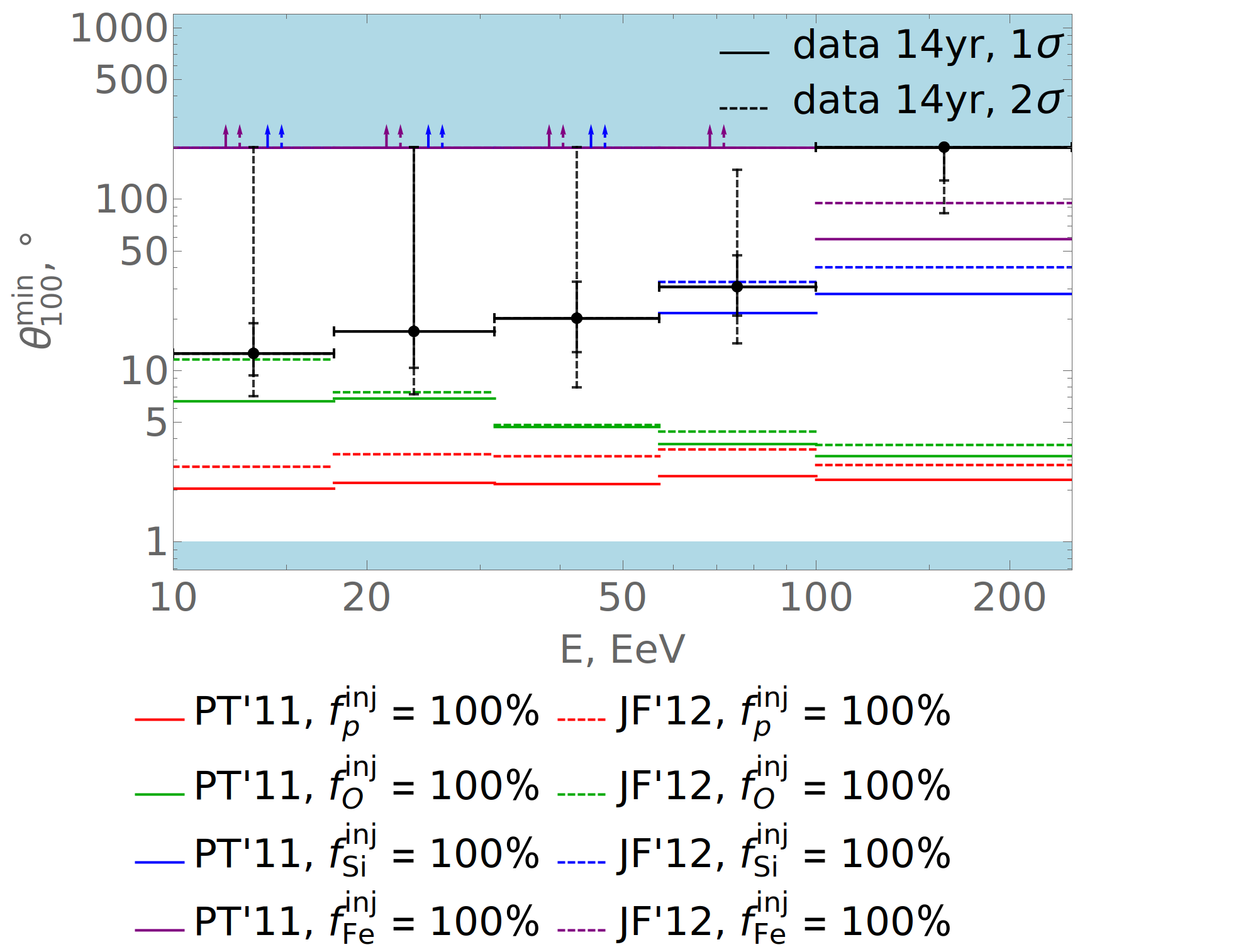}
 \includegraphics[width=0.99\columnwidth]{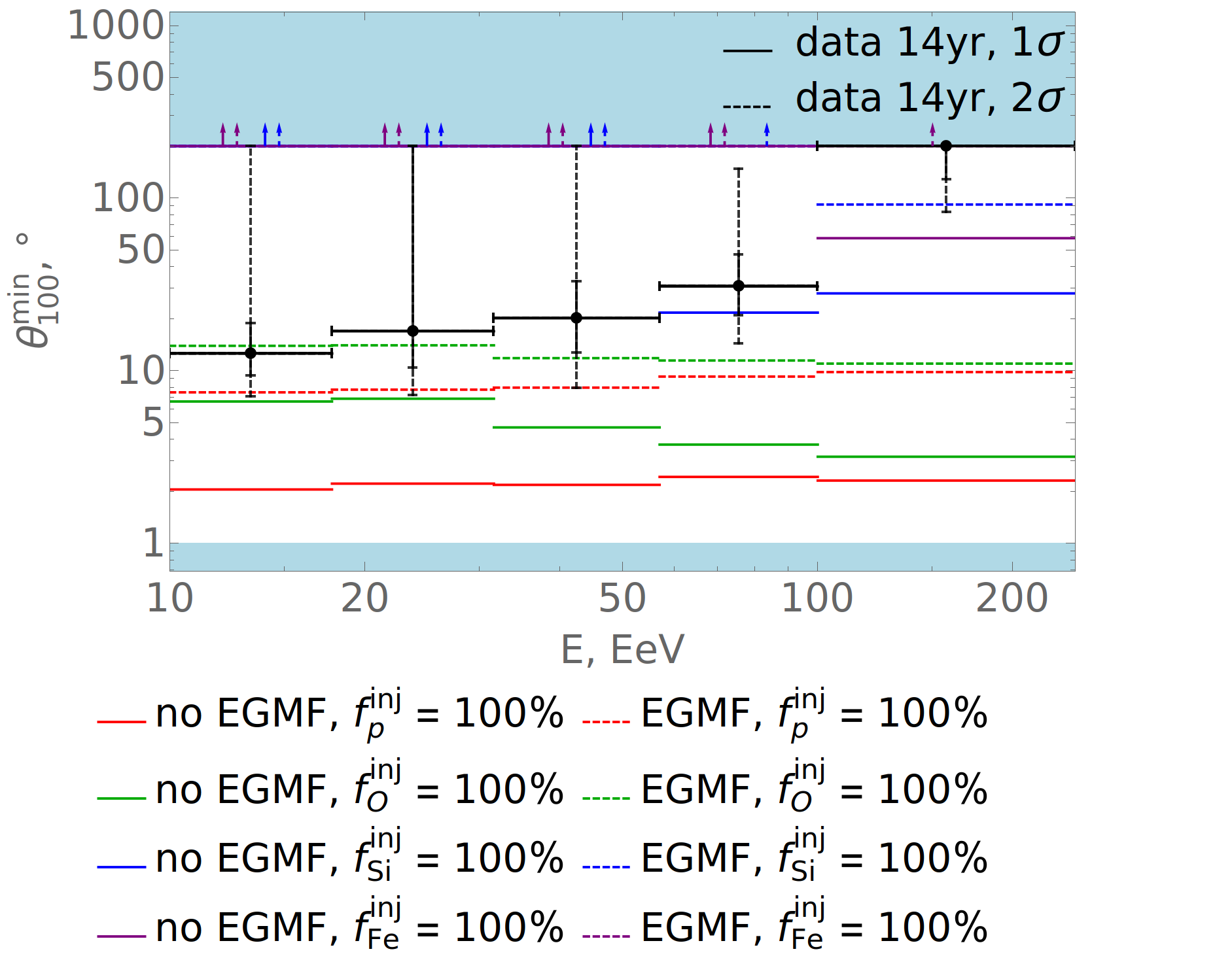}
\caption{
The test statistics for the data compared to various pure nuclei injected
composition models.
{\it Left panel:} results for two different regular GMF models.
{\it Right panel:} results without EGMF and with extremely strong EGMF.}
\label{fig:uncert-MF}
\end{figure*}

\begin{figure*}
 \includegraphics[width=0.99\columnwidth]{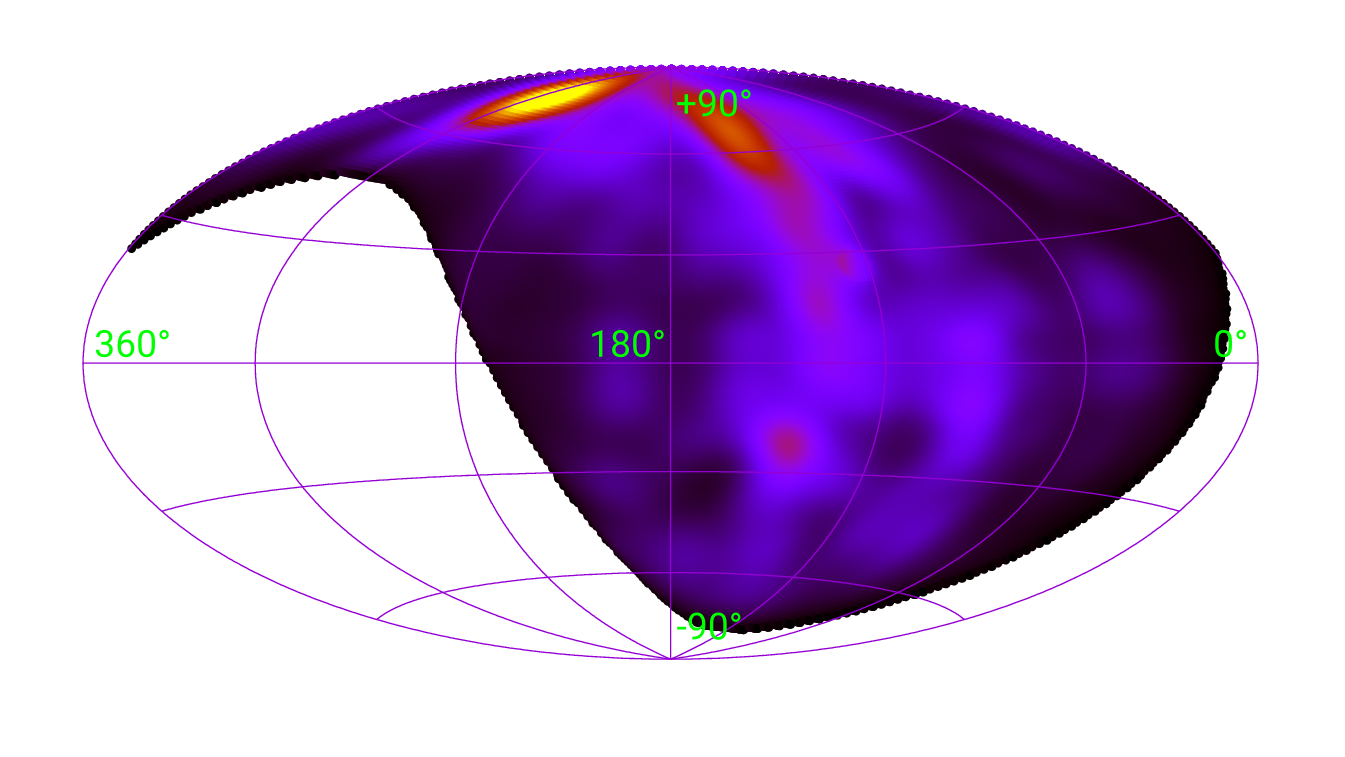}
 \includegraphics[width=0.99\columnwidth]{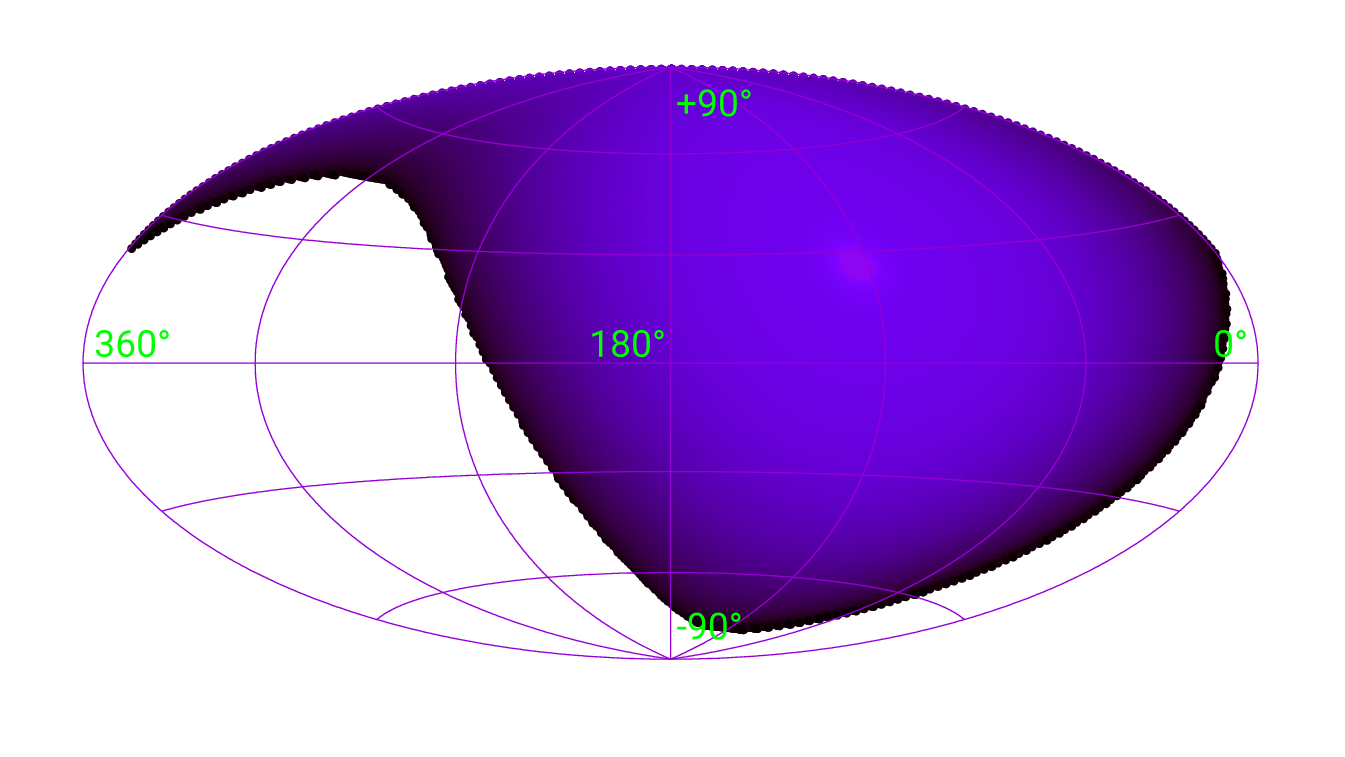}
\caption{
\label{fig:maps_EGMF}
Examples of UHECR flux model maps used for mock UHECR sets simulation. Deflections in ``extreme EGMF'' are assumed along with regular GMF model of Ref.~\cite{Pshirkov:2011um} and b-dependent smearing in random GMF model of Ref.~\cite{Pshirkov:2013wka}.
{\it Left:} protons at 100~EeV.
{\it Right:} iron at 100~EeV.
Maps are shown in galactic coordinates for TA SD field of view.
}
\end{figure*}

The extra-galactic magnetic field is much more uncertain than the Galactic one.
Only loose bounds on EGMF strength in voids were so far derived from observations: the constraint by $\sim 10^{-15}$~G from below~\cite{Neronov:1900zz, Taylor:2011bn}
and by $\sim 10^{-9}$~G from above~\cite{Pshirkov:2015tua}. The correlation length of a field of non-cosmological origin should not be larger than 1~Mpc~\cite{Durrer:2013pga}. There are also constraints from CMB on a field of cosmological origin with unbounded  correlation length: its strength should not exceed $5 \times 10^{-11}$~G~\cite{Jedamzik:2018itu}.
The contribution of the EGMF in the voids to the UHECR deflections is the largest for the field strength at its upper-limit value $B = 1.7$~nG and maximum correlation length of $\lambda = 1$~Mpc.
In this case the deflections for the protons at 100~EeV are as large as $7^\circ$ (we assume a distance traveled to be 250~Mpc --- the limit of our source catalog).
Note that this estimation is a conservative upper bound, as this deflection 
is assumed to be the same for all sources irrespective of their distance from us. Moreover, the deflection
is computed for the {\it detected} energy of the particle, while in reality it is accumulated during the whole path of the particle while its energy is higher.
We call this scenario ``extreme EGMF'' and model it with a uniform smearing of the catalog sources (according to particle charge and energy) before applying the deflections in GMF. Examples of UHECR flux model maps used for mock UHECR sets simulation for protons and iron nuclei in extreme EGMF scenario are shown in Fig.~\ref{fig:maps_EGMF}.

Apart from global EGMF in voids there can exist a magnetic field inside the extragalactic structures such as filaments. These fields require separate consideration in case our Galaxy itself is situated in a magnetized filament. 
While upper limits exist on the magnetic field strength of filaments in general~\cite{Brown:2017dwx, Locatelli:2021byc}, even a presence of such a structure around the Milky Way is unclear from observations, not to say of its magnetic field properties.
Therefore, for possible estimation of these fields it is reasonable to resort to results of the structure formation simulations~\cite{Hackstein:2016pwa, Hackstein:2017pex, Garcia:2020kxm}. 
For instance, the recent constrained simulation of EGMF in the local Universe~\cite{Hackstein:2017pex} shows the presence of a $\simeq 5$~Mpc-large local filament around the Milky Way magnetized to $\sim 0.3 - 3$~nG over most of its volume in a most conservative case.
The impact of this field on UHECR deflections would be smaller than that of our ``extreme EGMF'' scenario even if its correlation length equals the size of filament,  $\lambda \simeq 5$~Mpc. Therefore, we consider the ``extreme EGMF'' scenario as the most conservative one in terms of deflections.

\begin{figure*}
 \includegraphics[width=0.99\columnwidth]{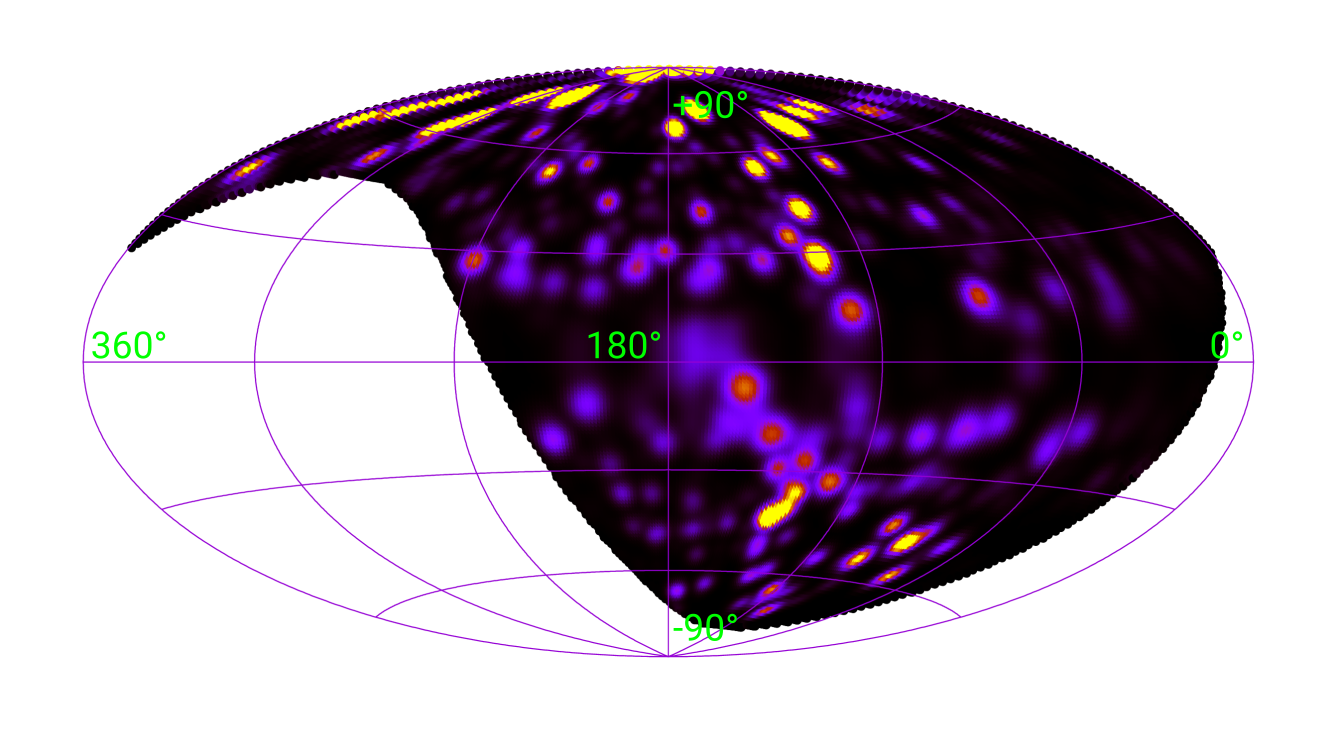}
 \includegraphics[width=0.99\columnwidth]{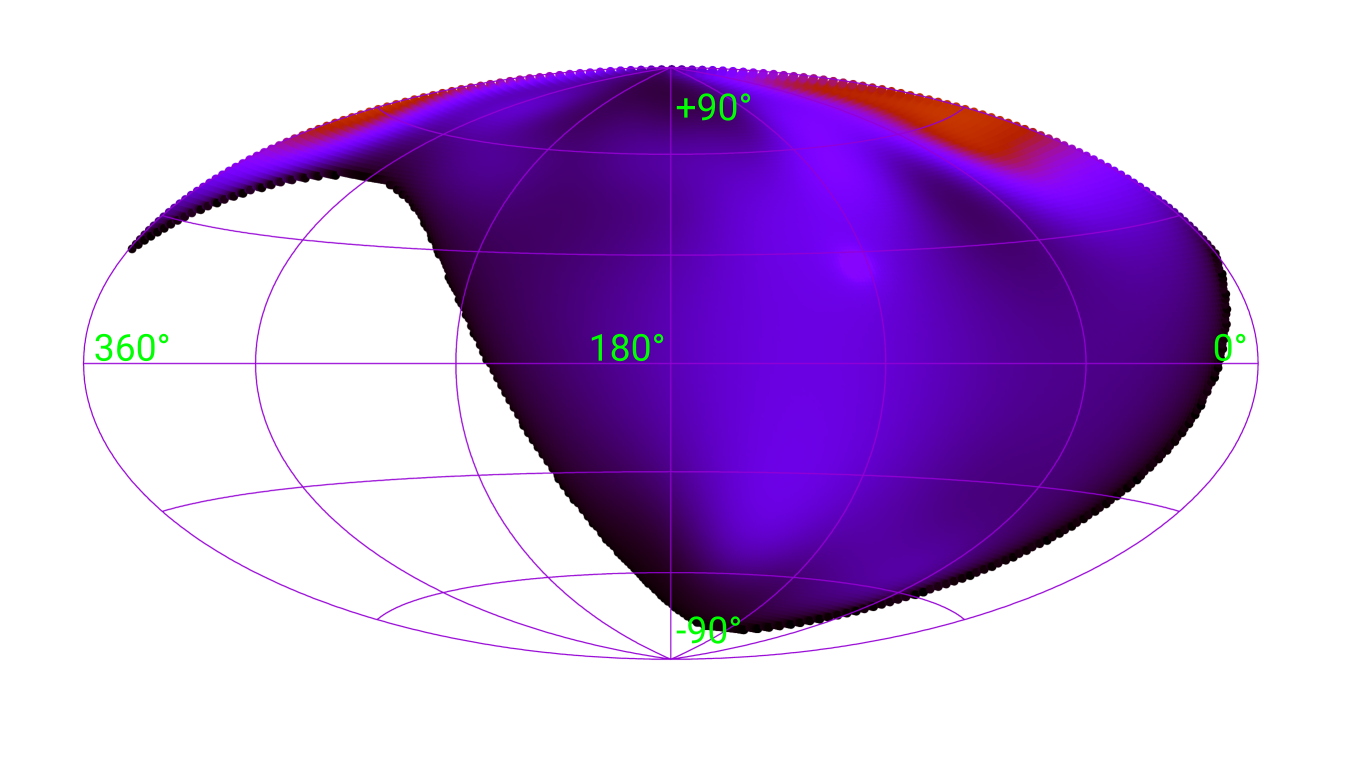}
 \includegraphics[width=0.99\columnwidth]{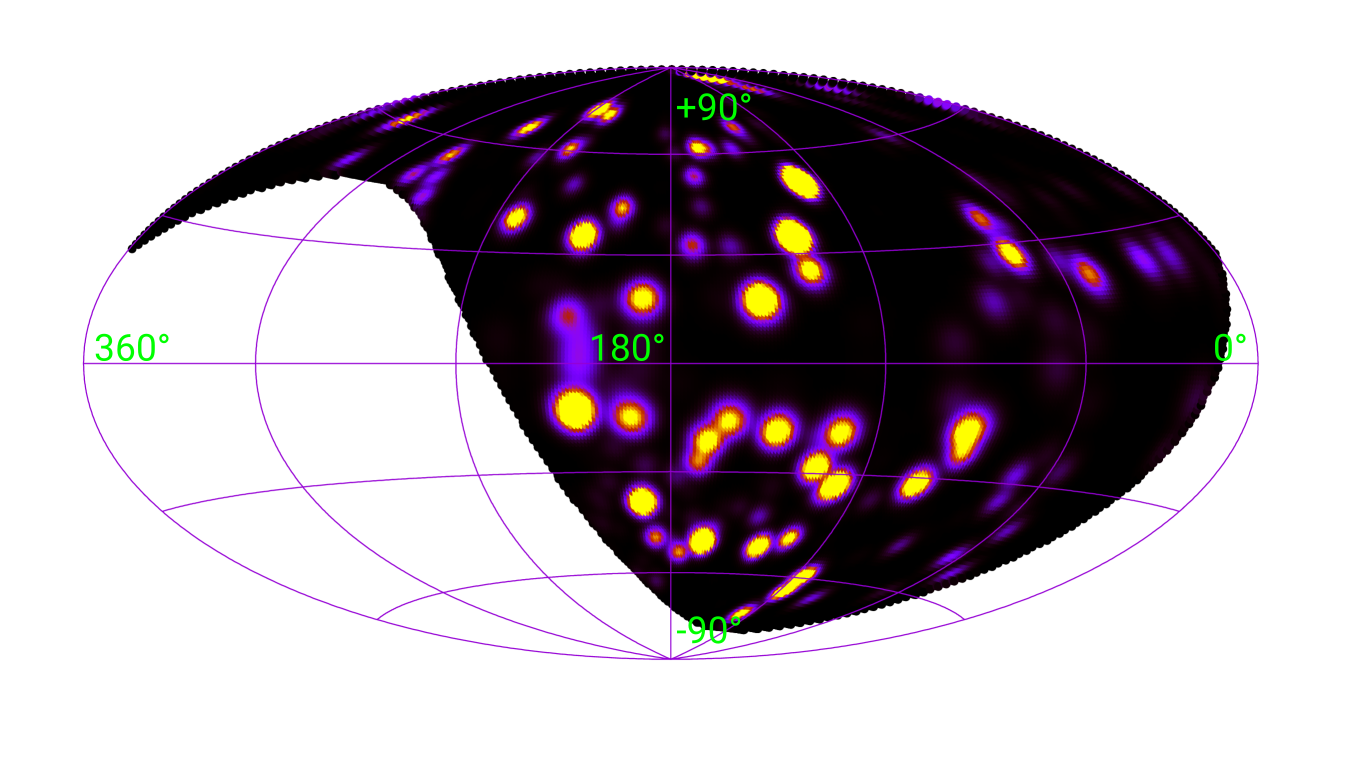}
 \includegraphics[width=0.99\columnwidth]{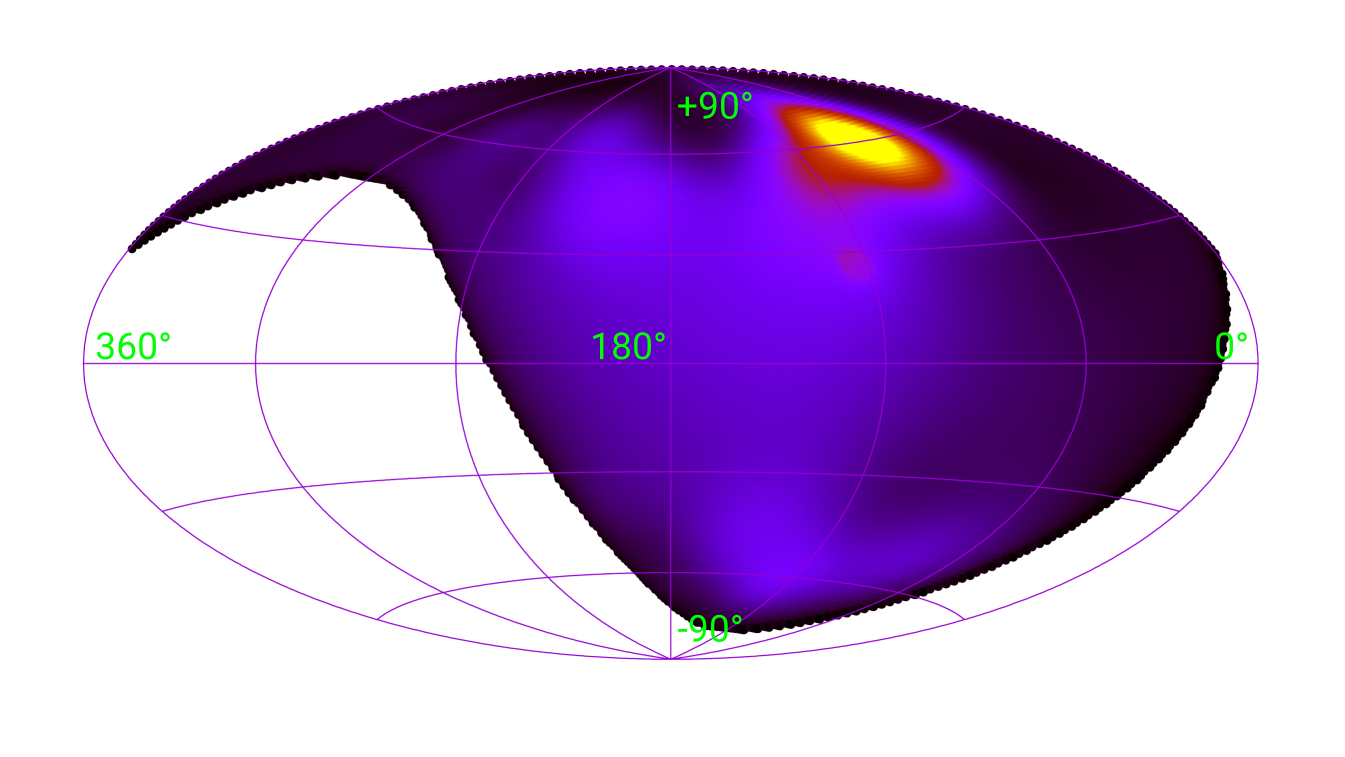}
\caption{
\label{fig:maps_rho}
Examples of UHECR flux model maps for mock sets simulated with rare sources. Deflections in regular GMF model of Ref.~\cite{Pshirkov:2011um} and b-dependent smearing in random GMF model of Ref.~\cite{Pshirkov:2013wka} are assumed. No EGMF deflections.
{\it Left column:} protons at 100~EeV for $\rho = 10^{-4}$~Mpc$^{-3}$ (top) and $\rho = 2 \times 10^{-5}$~Mpc$^{-3}$ (bottom).
{\it Right column:} iron at 100~EeV for $\rho = 10^{-4}$~Mpc$^{-3}$ (top) and $\rho = 2 \times 10^{-5}$~Mpc$^{-3}$ (bottom).
Maps are shown in galactic coordinates for TA SD field of view.
}
\end{figure*}

\begin{figure*}
 \includegraphics[width=0.99\columnwidth]{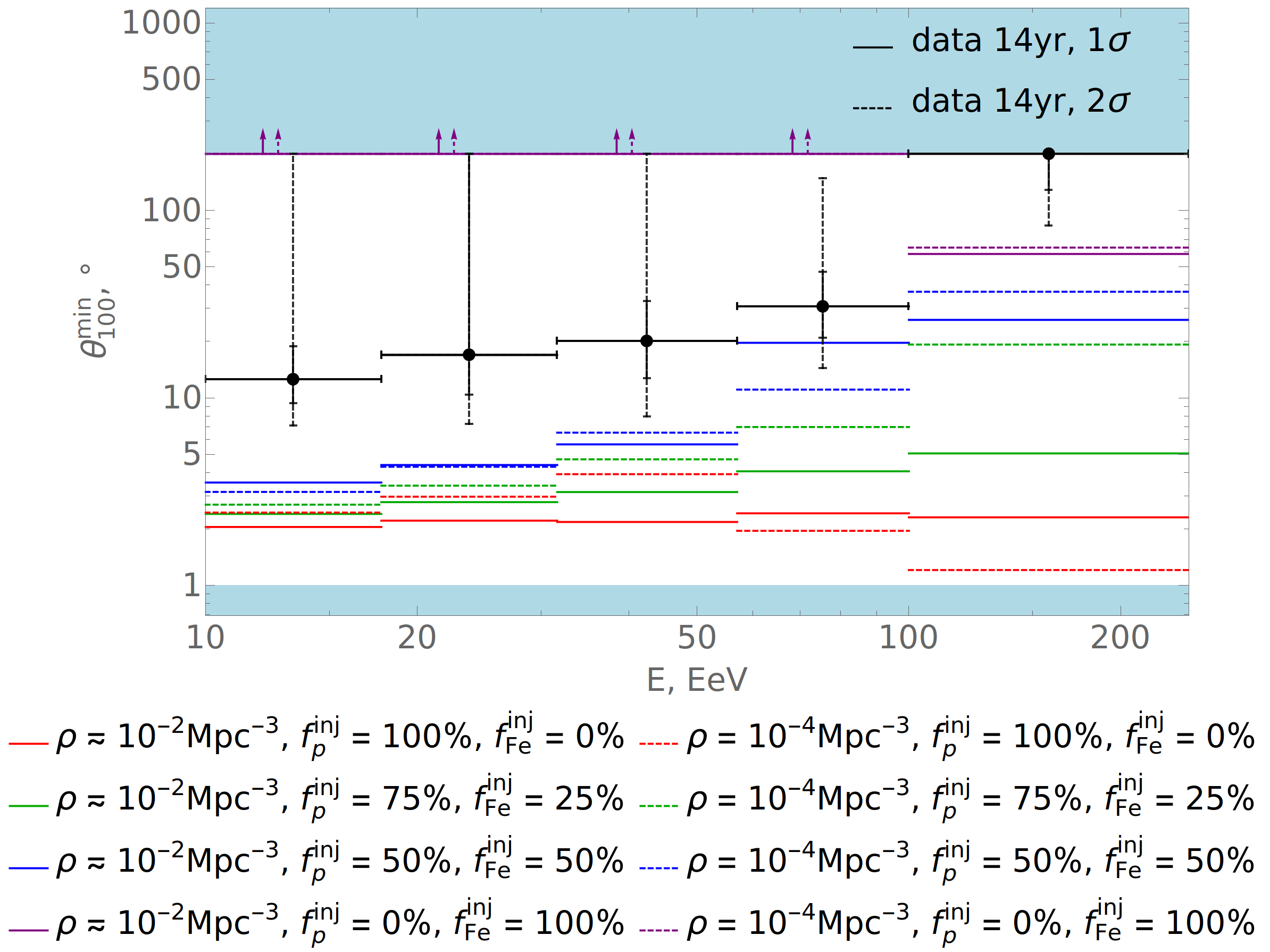}
 \includegraphics[width=0.99\columnwidth]{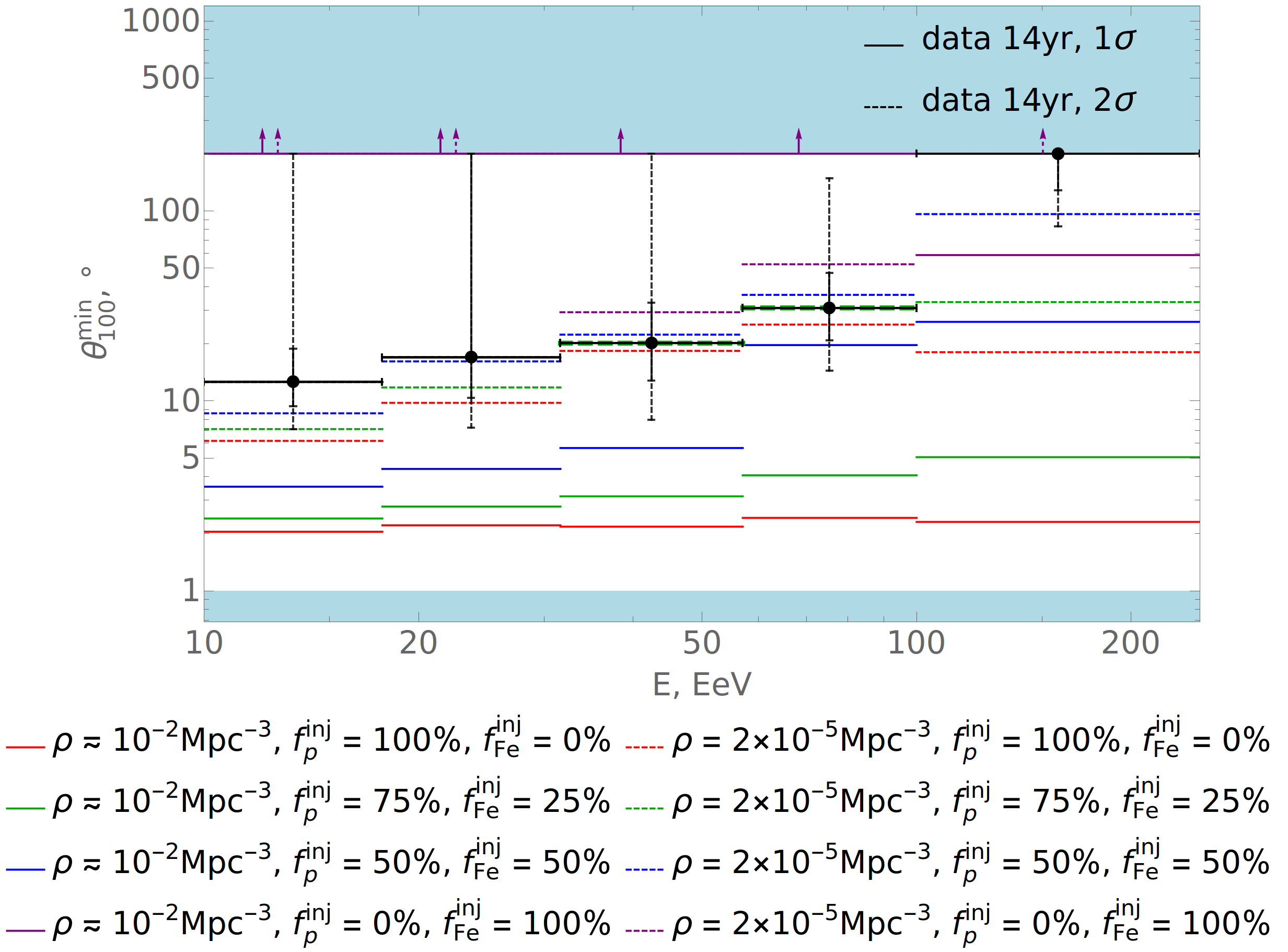}
\caption{
The test statistics for the data compared to reference composition models
for all-LSS sources and for rare sources in the LSS. The particular realizations of sources catalogs are the same as plotted in Fig.~\ref{fig:maps_rho}.
{\it Left panel:} p-Fe mix composition models for all-LSS sources and sources with number density $\rho = 10^{-4}$~Mpc$^{-3}$.
{\it Right panel:} p-Fe mix composition models for all-LSS sources and sources with number density $\rho = 2 \times 10^{-5}$~Mpc$^{-3}$.}
\label{fig:uncert-rho}
\end{figure*}

In Fig.~\ref{fig:uncert-MF} we show the comparison of the predicted TS for the same composition model computed without EGMF deflections and with deflections in the extreme EGMF scenario. One can see that the presence of EGMF affects the model predictions significantly,  so that even pure proton composition becomes consistent with data at $\sim 2 \sigma$ level at low energies. However, this does not hold at energies $E > 100$~EeV where all compositions lighter than silicon are still inconsistent with the data.
To reconcile the proton or helium composition models with the data at $E > 100$~EeV at least at $2\sigma$ level the EGMF should be stronger than 20~nG for $\lambda = 1$~Mpc, that is far beyond the upper limit discussed earlier in this section. We also should stress that this conclusion is conservative, since our procedure of estimation of EGMF deflections, that was described in the beginning of this subsection, definitely overestimates it. 

\subsection{Sources number density}
\label{sec:uncert_sources}
The largest uncertainty of composition models in our method is related to UHECR source number density.
As it was described earlier, the TS is computed assuming the conservative model where all the galaxies are equally luminous sources of UHECR.  The source number density is thus $\rho \simeq 10^{-2}$~Mpc$^{-3}$~\cite{binney2008galactic}.
However, the UHECR sources may be much more rare than ordinary galaxies.
The constraints on the source number density were placed by the Pierre Auger observatory in Ref.~\cite{PierreAuger:2013waq}.
For the scenario of sources in the LSS and at energies higher then 80~EeV the conservative 95\%~C.L. constraint is $\rho > 2 \times 10^{-5}$~Mpc$^{-3}$. 
However, this bound assumes the deflection of events not larger than $30^\circ$, that does not cover scenarios with heavy nuclei even in the case of deflections in GMF only.
There are two recent studies that are placing more stringent lower limits on the UHECR source number density: $\rho > 1.0 \times 10^{-4}$~Mpc$^{-3}$~\cite{Kuznetsov:2023jfw} and $\rho \gtrsim 3 \times 10^{-4}$~Mpc$^{-3}$~\cite{Bister:2023icg}. However, in the first of these works the density is constrained only for sources emitting heavy particles, while in the second one the constraints are put at energies $E \simeq 32$~EeV, while the sources at higher energies can be more rare.
At the same time, the viable UHECR sources being discussed recently include FR-I and Seyfert galaxies with $\rho \geq 10^{-4}$~Mpc$^{-3}$ in both cases, or even an order of magnitude more frequent low-luminosity AGNs~\cite{Kachelriess:2022phl}.

To test the robustness of the TS predictions to source number density we keep the source catalog for the TS computation fixed to our basic one and vary the catalogs used for mock event sets generation, while keeping all other model parameters fixed.
Namely, we test the conservative value from the Auger constraints: $\rho = 2 \times 10^{-5}$~Mpc$^{-3}$ and the benchmark value $\rho = 10^{-4}$~Mpc$^{-3}$.
We do not want to tie ourselves to any specific source class model, therefore we produce the test rare source catalogs from our basic all-galaxies catalog. 
For such low source number densities only a few or a few tens of source can be found in the local Universe and hence in the GZK sphere.
Therefore, the expected flux map starts to depend on the particular positions of these sources in the sky.
To avoid this statistical issue we generate a number of mock source catalogs and compute the TS for each of them separately.
The catalogs are volume limited samples generated by random selection from the original 2MRS catalog.
We generate 20 catalogs for both $\rho = 10^{-4}$~Mpc$^{-3}$ and $\rho = 2 \times 10^{-5}$~Mpc$^{-3}$ scenarios to keep the accuracy of the conclusion at the $95 \%$ level.

Among mock catalog realizations, in both $\rho = 10^{-4}$~Mpc$^{-3}$ and $\rho = 2 \times 10^{-5}$~Mpc$^{-3}$ cases we pick the catalog that gives the results that are most discrepant from that of the basic 2MRS catalog. The examples of respective UHECR flux model maps used for mock UHECR sets simulation for protons and iron nuclei are shown in Fig.~\ref{fig:maps_rho}.
The results are shown in Fig.~\ref{fig:uncert-rho}.
One can see that the discrepancy in TS between the basic scenario and the one with $\rho = 10^{-4}$~Mpc$^{-3}$ is not very large and does not exceed the difference between light and heavy composition models.
Therefore almost all of the conclusions that can be made for the basic source model stay in force.
The discrepancy between the basic scenario and the one with $\rho = 2 \times 10^{-5}$~Mpc$^{-3}$ is more pronounced, so that the light and intermediate compositions are mostly consistent with the data in the lower energy bins.
However, in the highest energy bin, the heavy composition is still preferred, while the light and intermediate compositions are in tension with the data.
We conclude that our method of composition estimation is robust to all the considered uncertainties, at least at highest energies.

At the same time, we suppose that one of the reasons of the
degradation of the TS sensitivity in the case of rare sources
is partial sky coverage of the TA experiment. For instance, at high energies and large deflections the sources are rare and some of them could contribute to the expected UHECR flux, while being outside the TA field of view. We have tested that in this situation the TS model separation power degrades. Conversely, for a full sky coverage even images of very smeared sources are fully inside the FoV and the TS separation power is higher. Therefore we expect that the sensitivity of our method would improve
if we use the UHECR data from the full sky, for instance by combining TA and Pierre Auger data in the style of Auger-TA Anisotropy Working Group studies~\cite{TelescopeArray:2021ygq, TelescopeArray:2021gxg}.

\section{Conclusion}
\label{sec:discussion}
\begin{figure*}[htbp!]
 \includegraphics[width=0.99\columnwidth]{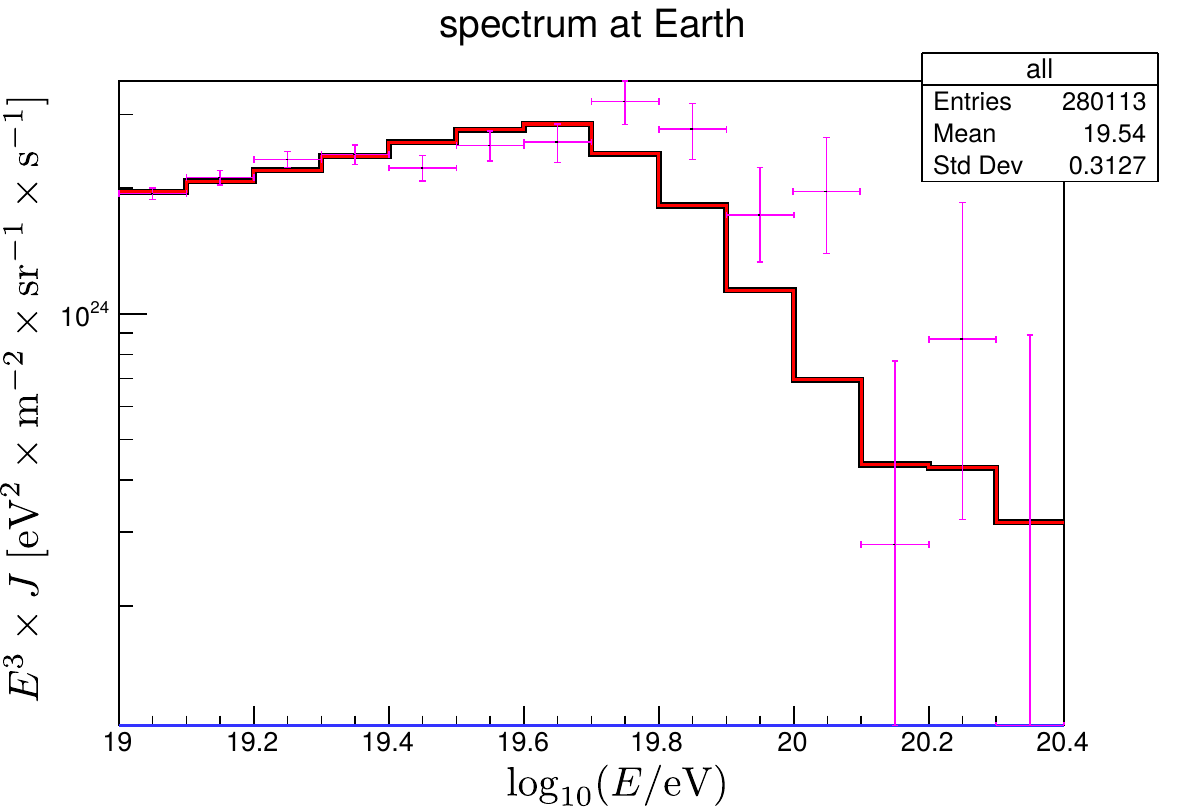}
 \includegraphics[width=0.99\columnwidth]{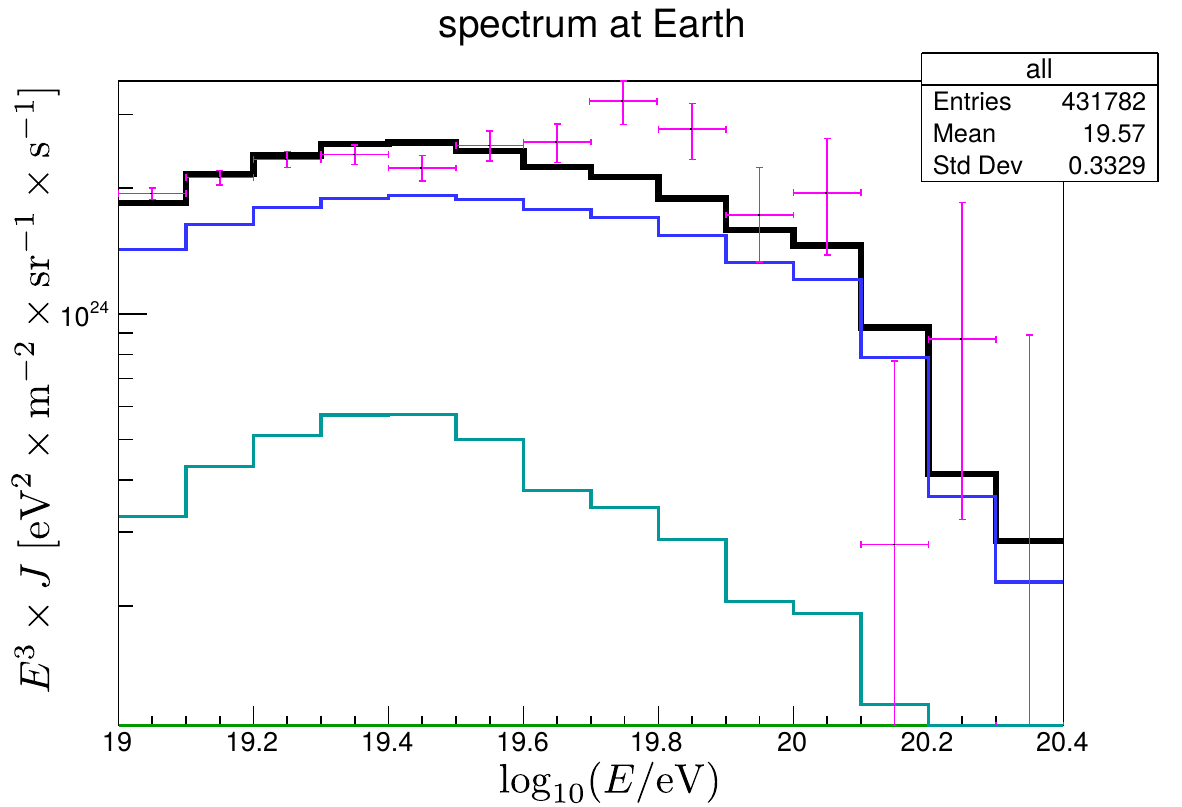}
\caption{
Fit of the observed TA spectrum with the pure p and pure Fe injected spectra. Red lines are observed protons, olive --- He, green --- Li to Ne, teal --- Na to Ar, blue --- K to Fe, black --- total, magenta crosses are data points. Some of the lines are merged together at zero.
{\it Left panel:} pure p injection without cutoff.
{\it Right panel:} pure Fe injection with a sharp cutoff at 560~EeV.
}
\label{fig:spectra_fits}
\end{figure*}

In this paper we have used the novel method proposed in Ref.~\cite{Kuznetsov:2020hso} to 
estimate the UHECR injected mass composition from the distribution of their arrival directions.
We improved the original version of the method
by attributing all the statistical uncertainties to the data and not to the composition models and therefore making the comparison of the models more transparent. We also applied the developed method to the Telescope Array SD data.

We tested several injected compositions: pure protons, helium, oxygen, silicon and iron, as well as a proton-iron mix in different proportions. For each model we propagated the injected particles taking into account the effects of their attenuation, production of secondary species, deflection in magnetic fields and modulation with the detector exposure. We then compared, separately in 5 energy ranges above 10~EeV, the resulting sky distributions of the mock events with the actual TA SD data. To assess quantitatively the compatibility of a given model with the data we calculated for both the test statistics (\ref{eq:TS}) as a function of the angle, and compared the positions of the minima,
which represent typical deflection angle of UHECR in a given set with respect to their sources in the LSS. The results presented in Fig.~\ref{fig:rails} indicate large deflections of UHECR, significantly larger than would normally be expected for a light composition. 

The main result of the present paper is the
thorough investigation of the stability of the new composition results with respect to all possible uncertainties: injected spectra, experiment energy scale, galactic and extragalactic magnetic fields and source number density.
We found that the preference for heavy composition is robust to the first three of these uncertainties at all energies. 
In the presence of a large extragalactic magnetic field or for very rare sources
light composition becomes marginally compatible with the data, but at highest energies the composition should be heavy in both of these cases.
We discuss the physical implications of the latter result in the short letter accompanying this study~\cite{TelescopeArray:2024oux}.

\appendix*
\section{Fits for injected spectra}
The fits for the injected spectra for protons and iron nuclei are shown in Fig.~\ref{fig:spectra_fits}. The TA data from Ref.~\cite{Ivanov:2020rqn} was used for these fits. For iron the fit for the injection spectrum with cutoff at 560~EeV is shown. The respective $\chi^2/{\rm d.o.f.}$ values are 1.80 for protons and 2.01 for iron nuclei.

\input{TAacknowledgements-20220308.tex}

\bibliography{ref.bib}

\end{document}

%% file: TA-author-20220427-revtex.tex
\author{R.U. Abbasi}
\affiliation{Department of Physics, Loyola University Chicago, Chicago, Illinois 60660, USA}

\author{Y. Abe}
\affiliation{Academic Assembly School of Science and Technology Institute of Engineering, Shinshu University, Nagano, Nagano 380-8554, Japan}

\author{T. Abu-Zayyad}
\affiliation{Department of Physics, Loyola University Chicago, Chicago, Illinois 60660, USA}
\affiliation{High Energy Astrophysics Institute and Department of Physics and Astronomy, University of Utah, Salt Lake City, Utah 84112-0830, USA}

\author{M. Allen}
\affiliation{High Energy Astrophysics Institute and Department of Physics and Astronomy, University of Utah, Salt Lake City, Utah 84112-0830, USA}

\author{Y. Arai}
\affiliation{Graduate School of Science, Osaka Metropolitan University, Sugimoto, Sumiyoshi, Osaka 558-8585, Japan}

\author{R. Arimura}
\affiliation{Graduate School of Science, Osaka Metropolitan University, Sugimoto, Sumiyoshi, Osaka 558-8585, Japan}

\author{E. Barcikowski}
\affiliation{High Energy Astrophysics Institute and Department of Physics and Astronomy, University of Utah, Salt Lake City, Utah 84112-0830, USA}

\author{J.W. Belz}
\affiliation{High Energy Astrophysics Institute and Department of Physics and Astronomy, University of Utah, Salt Lake City, Utah 84112-0830, USA}

\author{D.R. Bergman}
\affiliation{High Energy Astrophysics Institute and Department of Physics and Astronomy, University of Utah, Salt Lake City, Utah 84112-0830, USA}

\author{S.A. Blake}
\affiliation{High Energy Astrophysics Institute and Department of Physics and Astronomy, University of Utah, Salt Lake City, Utah 84112-0830, USA}

% ...

\collaboration{The Telescope Array Collaboration}
\noaffiliation

%% file: TAacknowledgements-20220308.tex
\section*{Acknowledgements}
The authors would like to thank the former member of the Telescope Array collaboration Armando di Matteo, who kindly provided the simulations of UHECR propagation and respective fits of attenuation curves for the purposes of this study.

The Telescope Array experiment is supported by the Japan Society for
the Promotion of Science(JSPS) through
Grants-in-Aid
for Priority Area
431,
for Specially Promoted Research
JP21000002,
for Scientific  Research (S)
JP19104006,
for Specially Promoted Research
JP15H05693,
for Scientific  Research (S)
JP19H05607,
for Scientific  Research (S)
JP15H05741,
for Science Research (A)
JP18H03705,
for Young Scientists (A)
JPH26707011,
and for Fostering Joint International Research (B)
JP19KK0074,
by the joint research program of the Institute for Cosmic Ray Research (ICRR), The University of Tokyo;
by the Pioneering Program of RIKEN for the Evolution of Matter in the Universe (r-EMU);
by the U.S. National Science
Foundation awards PHY-1806797, PHY-2012934, and PHY-2112904, PHY-2209583, PHY-2209584, and PHY-2310163, as well as AGS-1613260, AGS-1844306, and AGS-2112709;
by the National Research Foundation of Korea
(2017K1A4A3015188, 2020R1A2C1008230, \& 2020R1A2C2102800) ;
by the Ministry of Science and Higher Education of the Russian Federation under the contract 075-15-2024-541, IISN project No. 4.4501.18 by the Belgian Science Policy under IUAP VII/37 (ULB), by the European Union and Czech Ministry of Education, Youth and Sports through the FORTE project No. CZ.02.01.01/00/22\_008/0004632, and by the Simons Foundation (00001470, NG). This work was partially supported by the grants of The joint research program of the Institute for Space-Earth Environmental Research, Nagoya University and Inter-University Research Program of the Institute for Cosmic Ray Research of University of Tokyo. The foundations of Dr. Ezekiel R. and Edna Wattis Dumke, Willard L. Eccles, and George S. and Dolores Dor\'e Eccles all helped with generous donations. The State of Utah supported the project through its Economic Development Board, and the University of Utah through the Office of the Vice President for Research. The experimental site became available through the cooperation of the Utah School and Institutional Trust Lands Administration (SITLA), U.S. Bureau of Land Management (BLM), and the U.S. Air Force. We appreciate the assistance of the State of Utah and Fillmore offices of the BLM in crafting the Plan of Development for the site.  We thank Patrick A.~Shea who assisted the collaboration with valuable advice and supported the collaboration’s efforts. The people and the officials of Millard County, Utah have been a source of steadfast and warm support for our work which we greatly appreciate. We are indebted to the Millard County Road Department for their efforts to maintain and clear the roads which get us to our sites. We gratefully acknowledge the contribution from the technical staffs of our home institutions. An allocation of computing resources from the Center for High Performance Computing at the University of Utah as well as the Academia Sinica Grid Computing Center (ASGC) is gratefully acknowledged.